\documentclass[aps,pre,showpacs,amsmath,amssymb,reprint]{revtex4-1}
\usepackage{amsmath}
\usepackage{amsfonts}
\usepackage{amsthm}
\usepackage{graphics}
\usepackage{graphicx}
\usepackage{subfigure}
\usepackage{amssymb}
\usepackage{color}
\usepackage{url}
\usepackage[abs]{overpic}
\usepackage[usenames,dvipsnames]{xcolor}
\usepackage[normalem]{ulem}

\def\r{\textbf{r}}

\graphicspath{{./}}

\def\R1{\tt{Run1}}
\def\R2{\tt{Run2}}
\def\R3{\tt{Run3}}

\begin{document}

\title{Non-equilibrium Bose-Einstein Condensation}

\author{Vishwanath Shukla}
\email{research.vishwanath@gmail.com}
\affiliation{Department of Physics, Indian Institute of Technology Kharagpur, Kharagpur - 721 302, India}%
\author{Sergey Nazarenko}%
\affiliation{Universit\'e C\^ote d'Azur, CNRS, Institut de Physique de Nice, Parc Valrose, 06108 Nice, France}%

\date{\today}
\begin{abstract} 
We investigate  formation of Bose-Einstein condensates under non-equilibrium conditions using numerical simulations of the three-dimensional Gross-Pitaevskii equation. For this, we set initial  random weakly nonlinear excitations and the forcing  at high wave numbers, and study propagation of the turbulent spectrum toward the low wave numbers.  Our primary goal is to compare the results for the evolving spectrum 
with the previous results obtained for the kinetic equation of weak wave turbulence. We demonstrate existence of a regime for which good agreement with the wave turbulence results is found in terms of the main features of the previously discussed self-similar solution. In particular, we find a reasonable agreement with the low-frequency and the high-frequency power-law asymptotics of the evolving solution, including the anomalous power-law exponent $x^* \approx 1.24$ for the three-dimensional waveaction spectrum. We also study the regimes of very weak turbulence, when the evolution is affected by the discreteness of the Fourier space, and the strong turbulence regime when emerging condensate modifies the wave dynamics and leads to formation of strongly nonlinear filamentary vortices.
\end{abstract}

\maketitle

\section{Introduction}
Bose-Einstein condensation (BEC), a macroscopic quantum phenomenon, is now routinely realized in laboratory experiments in systems involving ultra cold atoms, polaritons-excitons, photons, magnons, etc. Moreover, BECs are suggested to play an important role in the physics of astrophysical objects such as neutron stars~\cite{migdal1959superfluidity,ginzburg1964superfluidity,anderson1975pulsar,shternin2011cooling,page2011rapid,haskell2015models} and in cosmology~\cite{kibble1976topology,zurek1985cosmological,hendry1994generation}. 
Formation of BEC under equilibrium settings and its properties have been extensively studied since the prediction of the phenomenon. Its connection with the phenomenon of superfluidity and the later experimental realizations in different systems makes it one of  the most fascinating, but challenging quantum many-body system with  many nontrivial macroscopic effects~\cite{anglin2002bose,proukakis2017universal}. 

Paths leading to a BEC need not be in equilibrium, it is well known that ultra-cold bose gases under cooling quenches pass through highly nonequilibrium states and may eventually lead to the formation of a condensate~\cite{ritter2007observing,beugnon2017exploring,gring2012relaxation,eigen2018universal,glidden2021bidirectional,nowak2012nonthermal,ranccon2014equilibrating,sadler2006spontaneous}. This nonequilibrium  relaxation dynamics of the quantum bose gases and other quantum many-body systems has attracted a major attention, still the full understanding and characterization of the ensuing dynamics is far from complete~\cite{hohenberg1977RMP,damle1996phase,lamacraft2007quantum,polkovnikov2011colloquium,gogolin2016equilibration,bhattacharyya2020universal,deng2018off,rodriguez2020turbulent,mitra2021hydrodynamic}. It has been argued that the dynamical evolution depends on the strength and nature of the cooling quench in closed systems
leading to either
wave turbulence dominated states~\cite{DNPZ,connaughton2005condensation,nazarenko2006wave,vmrnjp13,SNBKT},
or non-thermal fixed points with universal scaling laws~\cite{nowak2012nonthermal,prufer2018observation,erne2018universal,glidden2021bidirectional}.

Important insights into the dynamical evolution of such system have been obtained based on the Boltzmann-like kinetic description involving the particle momentum distribution function~\cite{levich1978time,snoke1989population,kagan1992erratum,svistunov1991highly,SemTka,SemTka1,ConPom}. It has been suggested that the system passes through three stages: in the first stage the distribution function undergoes a kinetic redistribution, essentially transporting particles in momentum space towards lower momenta; the second stage is characterized by the nucleation of a condensate; finally, the third stage involves growth of the condensate and its interaction with uncondensed particles. However, the kinetic approach is not applicable for the second stage, so it can not describe the development of coherence~\cite{svistunov1991highly,SemTka,ConPom}. 

A more general framework to study the nonlinear nonequilibirum dynamical evolution of either quenched or forced-dissipated systems of weakly interacting ultracold bose gases is the Gross-Pitaevskii equation (GPE) and its variants that provide a semi-classical description~\cite{berloff2002scenario,DNOQuantumcascade,nazarenko2006wave,proment2012sustained}. GPE is also referred to as the nonlinear Schr\"odinger equation (NLSE). GPE based theories can describe the dynamics of the uncondensed waves, formation of the condensate, and subsequently its interaction with the uncondensed component both in equilibrium and out of it~\cite{davis2001simulations,blakie2008dynamics,krstulovic2011energy,vmrnjp13,berloff2014modeling}. Moreover, it provides a quantitative description of weakly interacting BECs, whereas it applies only qualitatively if the boson-boson interactions are strong, for example in the case of superfluid helium. Therefore, it seems to be an appropriate choice to examine the post-quench dynamics of both closed and open systems. In particular, our emphasis in this work is on the forced-dissipated systems, which are useful for exciton-polariton condensates or certain astrophysical system such as Bose stars, to understand the universal features of the dynamics leading to the condensation and test them against the theoretical wave turbulence predictions and existing numerical simulations of the wave-kinetic equation (WKE).

As mentioned above, BEC may occur out of initial states that are far from thermal equilibrium. This situation is particularly interesting because the condensation process in such cases is strongly non-equilibrium and it shares fundamental features that are common for turbulent systems -- cascade-like transfer of excitations through the momentum space. Moreover, in the absence of forcing and dissipation, the system has two conserved quantities, the total energy and the total number of particles. Therefore, it has a dual-cascade property similar to the one in turbulence of classical two-dimensional flows that conserve the total energy and the total enstrophy~\cite{kraichnan2DRP,DNPZ,SNbook,VS2DHVBK2015}. This analogy allows us to interpret the non-equilibrium condensation process as an inverse cascade of mass in the momentum space (i.e. from large to low momenta), whereas the creation of high-momentum particles and their eventual escape from the system (``evaporation") as a direct energy cascade~\cite{Lvov2003,SNbook}. Of special interest is the universal scaling behaviour in the evolving turbulent spectrum that is anticipated and was previously discussed in literature in the context of the nonequilibrium condensation ~\cite{SemTka,SemTka1,LacPom,Semisalov2021}.

Our primary objective in this paper is to elucidate the dynamics leading to the formation of a BEC under nonequilibrium settings within the GPE framework in three-dimensions (3D). In particular,  we numerically solve the GPE in the inverse cascade setting to compare the obtained results with those of the WKE, including the pre-$t_{\star}$ self-similar behaviour and the post-$t_{\star}$ thermalisation process, where $t_{\star}$ denotes the finite time in which the WKE-described distribution function blows up to infinity. We also consider setups where the applicability conditions for the WKE are violated (initially or as a result of evolution) either because the nonlinearity is too weak, leading to the $k$-space discreteness effects, or too strong, leading to presence of coherent structures such as strong condensate and hydrodynamic vortices.

The remainder of this paper is organized as follows. In Sec.~\ref{sec:gpetheory} we give an overview of the GPE framework as appropriate for this work. Section~\ref{sec:WTP} provides a brief account of the wave turbulence theory, where focus is on the WKE predictions for the nonequilibrium BEC. In Sec.~\ref{sec:numerics} we summarize the numerical methods and the parameters that are used in this study. 
Section~\ref{sec:results} contains results of our numerical simulations and their discussion, while in Sec.~\ref{sec:conclusions} we give the conclusions drawn from results and discuss the significance of our work.

\section{Gross-Pitaevskii theory}
\label{sec:gpetheory}

In the present paper, we study the non-equilibrium condensation by computing the  spatio-temporal evolution of the complex, classical wave function $\psi(\mathbf{r},t)$ of a weakly interacting Bose gas  described by the Gross-Pitaevskii equation (GPE),
\begin{equation}\label{GPEalpha}
i\frac{\partial \psi}{\partial t} = -\alpha\nabla^2\psi + g|\psi|^2\psi 
+ i\,F(\mathbf{r},t) - i\,\nu_h(-\nabla^2)^8\psi,
\end{equation}
where  $\alpha\equiv\hbar/2m$, 
$m$ the mass of a constituent boson, and $\hbar$ the Planck constant, $g$ is the effective interaction strength. To investigate the non-equilibrium Bose-Einstein condensation, we introduce a forcing that excites the system at small length scales. The explicit form of the forcing term $F$, written in Fourier space, reads $\widehat{F}=f_0\exp{(i\Theta(\mathbf{k},t))}$, wherein $f_0$ is the  forcing amplitude, which is finite within a wave number shell $k_{f1} \le |\mathbf{k}|\le k_{f2}$ and zero otherwise, and $\Theta(\mathbf{k},t)$ represents the random (forcing) phase which is uniformly distributed in $(0,2\pi]$ and independent at each time step and at each wave number. Our attention will be to the range of wave numbers between  $k=0$, the mode at (and near) which the condensation occurs, and the forcing wave numbers.  Moreover, in this study we do not focus on the forward cascade process from the forcing wave numbers to higher wave numbers, as the cost of resolving both the cascades in a numerical simulation is prohibitively expensive. 

In order to achieve a large range of wave numbers for the inverse cascade, the forcing range boundaries  $k_{f1} $ and $k_{f2}$ are shifted to higher wave numbers followed by introduction of a dissipation that acts at still higher wave numbers, effectively killing the forward cascade process. This dissipation mechanism is modeled by using a hyperviscosity represented by the last term in  Eq.~\eqref{GPEalpha} (with hyperviscosity coefficient $\nu_h =$~const), which is often used in numerical simulations of turbulent flows; see e.g.~\cite{DNOQuantumcascade}. If used carefully, hyperviscosity helps in obtaining a large wave number range for a cascade process, while avoiding the effects of truncation (finite wave numbers). We do not introduce any dissipation mechanism at large scales (low wave numbers), where condensate is formed. This allows us to obtain non-stationary turbulent states with condensation and study the self-similar evolution that is not influenced by the dissipation. 

Note that our forced-dissipated system is not only useful for the theoretical and numerical studies, but it also models a recent forced-dissipated BEC experiment where the forcing was done via ``shaking" the retaining trap and the dissipation---via using a finite-height trap allowing energetic atoms to escape from the system~\cite{navon2016}.

In the absence of forcing and dissipation, the total mass and the total energy of the bosons are conserved, 
\begin{equation}
M \equiv \int_{\mathcal{V}} d\mathbf{r}\,{\rho(\mathbf{r},t)} = \hbox{const},  
\end{equation}
where $\rho(\mathbf{r},t)\equiv 
|\psi(\mathbf{r},t)|^2$ is the condensate mass density, and
\begin{equation}
E \equiv \frac{1}{2} \int_{\mathcal{V}} d\mathbf{r}\, \left(\alpha |\nabla \psi(\mathbf{r},t)|^2 + \frac{g} {2} {|\psi(\mathbf{r},t)|^4}\right) = \hbox{const}.
\end{equation}
Here, the integration is performed over the entire volume of the system assuming suitable boundary conditions. In the present study, $\mathcal{V} $ is a periodic cube of side $L$ in the 3D space. The energy functional serves as a Hamiltonian for the GPE (in absence of forcing and dissipation) with its quadratic part $E_2= \frac{\alpha}{2} \int d\mathbf{r}\, |\nabla \psi(\mathbf{r},t)|^2 $ and the quadric part  $E_4 = \frac{g}{4} \int d\mathbf{r}\, {\rho(\mathbf{r},t)^2} $ corresponding to the linear and the nonlinear terms respectively.

In the presence of forcing and dissipation, the total mass and energy vary with time.  Other useful quantities are the local speed of sound $c=\sqrt{2\alpha\,g\,\rho}
$, and the healing length $\xi=\sqrt{\alpha/g\rho}$.  The healing length quantifies the distance over which the condensate density recovers from being zero at infinitely hard confining walls to a bulk value; it also gives the radius of the superfluid vortex core.  Note that $4\pi\alpha = h/m$ is the quantum of circulation.

\section{Wave turbulence predictions}
\label{sec:WTP}

A strongly turbulent state of a 3D superfluid involves an interacting, dynamic tangle of quantized vortices on top of a condensate in the presence of random waves and other coherent structures. But strong turbulence is not the only possibility: nonlinearly interacting random waves give rise to an out-of-equilibrium system, called wave turbulence, characterized by the presence of inter-length-scale transfers (cascades) of certain (conserved) quantities.
An important role in understanding  turbulent BEC, and the processes leading to its formation, is played by the statistical description of random weakly nonlinear waves -- the so-called wave turbulence theory (WTT)~\cite{DNPZ,SNbook,SemTka,SemTka1,LacPom,ConPom,kolmakov2014wave}. 

The WTT starts by considering the wave system in a periodic box of side $L$, in the present case a cubic domain of volume $ \mathcal{V} =L^3$. The waves constituting the system are \emph{assumed} to have random independent phases and amplitudes (RPA) at each wave number. Moreover,  in WTT  the infinite box limit $L\to \infty$ is taken before the weak-nonlinearity limit, which means that there is a large number of quasi-resonances each of which is as important as the exact wave resonance (for a detailed discussion see Ref.~\cite{SNbook}). One of the important objects in WTT is the wave action spectrum. For our system, it is defined as 
\begin{equation}
n({\bf k},t) = \lim_{L\to \infty} \left[ \frac{L^3}{(2\pi)^3} \langle |\hat \psi ({\bf k},t) |^2  \rangle \right],
\end{equation}
where $\hat{\psi}$ is the Fourier coefficient,
\begin{equation}
\hat{\psi} ({\bf k},t) = \frac 1 {L^3} \int_{\mathcal{V}} d\mathbf{r}\, \psi ({\bf r},t) e^{-i  ({\bf k} \cdot {\bf x})  },
\end{equation}
and $\langle \cdot \rangle$ denotes the RPA averaging.

The WTT analysis of the GPE \eqref{GPEalpha}, in the absence of forcing and dissipation, along with the assumption of isotropy  leads to the following wave-kinetic  equation (WKE) for the wave action spectrum, 
\begin{eqnarray}\label{E1}
\!\!\!\!\!\!\!\!\!\!\!\!\!\!\!\!\!\!\!\!
\frac{\partial  n_{\omega} }{\partial t}= \pi \omega^{-1/2}\int  \, S(\omega,\omega_1,\omega_2,\omega_3) 
\delta(\omega + \omega_1 - \omega_2
- \omega_3)\\
n_{\omega}n_{1}n_{2}n_{3} \left(n_{\omega}^{-1} + n_{1}^{-1} - n_{2}^{-1} - n_{3}^{-1}\right) d\omega_1d\omega_2d\omega_3, \nonumber
\end{eqnarray}
 see e.g.  Ref.~\cite{DNPZ}. Observe that we have switched from the wave number to the frequency space: $n_{\omega}(t)=n(k(\omega),t)$, where $\omega(k)=k^2$ is the wave frequency, $k=|{\bf k}|$, and have introduced short-hand notations $n_i=n(\omega_i,t)$, $\omega_i=k_i^2$, $i=1,2,3$. The integral in   (\ref{E1}) is taken over the positive values of $\omega_{1}$, $\omega_{2}$ and $\omega_{3}$. The kernel of the integral is
\begin{equation}
S(\omega,\omega_1,\omega_2,\omega_3) = \min\left(\sqrt{\omega},\sqrt{\omega_1},\sqrt{\omega_2},\sqrt{\omega_3}\right).
\end{equation}
Like the original GPE \eqref{GPEalpha},    Eq. \eqref{E1} conserves the total mass and the total energy, which now look as follows,
\begin{equation}
M =  2\pi \int_0^\infty  \omega^{1/2} \, n_{\omega} \, d \omega,
\end{equation}
and 
\begin{equation}
E = 2 \pi \int_0^\infty   \omega^{3/2}  \, n_{\omega} \, d \omega.
\end{equation}

Equation~\eqref{E1} has two thermodynamic equilibrium solutions $n_{\omega} =$~const and $n_{\omega} \sim \omega^{-1}$ corresponding to the particle and the energy equipartition in the 3D $k$-space, respectively. It also has two non-equilibrium stationary power law spectra  $n_{\omega} \sim \omega^{-x}$ with $x = 7/6$ and $x = 3/2$. These  are the Kolmogorov--Zakharov (KZ) spectra corresponding to  the inverse and direct cascades of $M$ and $E$, respectively~\cite{DNPZ,SNbook}.

The non-stationary solutions of the  WKE~(\ref{E1})  in the condensation (inverse cascade) settings were studied numerically in~\cite{SemTka,SemTka1,LacPom,ConPom}. The spectrum was shown to have  tendency to blow up to infinity  in a finite time $t_{\star}$ at $\omega=0$ (corresponding to $k=0$). Moreover, these studies showed that a power-law behavior $n_{\omega}(t) \sim \omega^{-x_{\star}}$ starts to develop in the tail and invades the whole inverse cascade range as $t  \to t_{\star}$. 
Naively, one could think that this power law has exponent $7/6$, as in the inverse cascade KZ spectrum. This statement was indeed made in \cite{svistunov1991highly,kagan1992erratum}, but later proven wrong in ~\cite{SemTka,SemTka1,LacPom,ConPom} wherein the numerically observed exponent is in the range $x_{\star} \approx 1.23-1.24 $, which is clearly different from the  exponent $7/6 \approx 1.167$. 

The blowup behaviour of the solution was attributed  to its property of self-similarity of the second kind. According to the Zeldovich-Raizer classification scheme, a solution whose similarity properties cannot be fully determined from a conservation law (conservation of mass in our case) has the self-similarity of the second kind~\cite{ZR}. 
{This is because the self-similar part of the evolving solution contains only a tiny part of the total mass. (The respective spectrum is said to have a finite capacity.) As a consequence, one cannot find  the exponent $x_{\star}$ of the asymptotic tail exactly: one has to solve a ``nonlinear eigenvalue problem". Such a nonlinear eigenvalue problem was addressed directly in \cite{Semisalov2021} where the self-similar solution to the WKE was found with $\sim5$\% accuracy and with  $x_{\star} \approx 1.22$. In this paper, the low-frequency asymptotics of the spectrum was rigorously proven to be independent of the frequency---a property which was previously observed numerically in 
\cite{SemTka,SemTka1}.
As we see, both direct solution of WKE~(\ref{E1})  to find the evolving spectrum and  the solution of (\ref{E1})  restricted to the similarity ansatz
lead to mutually consistent results.}

Furthermore, it was argued in~\cite{SemTka,SemTka1}, and supported by numerical simulations of the Boltzmann kinetic equations, that the post-blowup evolution leads to the creation of a condensate and a thermalised component whose spectral signatures are the formation of a Dirac-delta spectrum at $k=0$ and an equiparition of energy for the $k\ne 0$ wave numbers, respectively.

Before concluding this section, below we briefly mention three facts from the WTT that will be required for the discussion of our results, for a detailed account of these see Ref.~\cite{SNbook}.

(A) The leading order nonlinear effect in the four-wave systems is actually not the spectrum evolution described by the WKE, but a nonlinear frequency shift $\Omega$ that leaves the spectrum unchanged,
\begin{equation}
\Omega = 2 g \langle \rho \rangle.
\label{Omega}
\end{equation}
This frequency shift will be seen in the spatio-temporal spectra measured in our simulations.

(B) Apart from the small nonlinearity, the applicability of the WKE requires that the finite-box effect and associated with it the discreteness of the $k$-space are not important. This means that the 
nonlinear frequency broadening  $\delta_\omega $ (which is of the order of inverse characteristic time of the spectrum evolution $\tau_{\tt WKE}$) must be larger than the distance 
between adjacent linear-wave frequencies $\Delta \omega \approx (d\omega_k/dk) \Delta k = 2 \alpha k (2\pi/L).$

(C) All of the WTT theory described above, including WKE~(\ref{E1}), refers to the four-wave regime characterised by a small nonlinearity,
$\eta=E_4/E_2 \ll 1$. However, if during the course of the evolution the condensate present at $k=0$ grows to a large value, then it starts to affect the dynamics of the uncondensed waves with $k\ne 0$. Moreover, if the condensate fraction $M_0/M \sim 1$, then such waves can be regarded as small perturbations $\tilde \psi$ over a strong uniform condensate state  $\psi_0$ (in the physical space), i.e. $|\tilde \psi| \ll |\psi_0|$. Under these circumstances, such waves acquire acoustic properties and their frequency satisfies the so-called Bogoliubov dispersion relation~\cite{bogoliubov1947theory}:
\begin{equation}
\omega_B(k) = g |\psi_0|^2 \pm \sqrt{2g |\psi_0|^2 k^2 + \alpha k^4}.
\label{bog}
\end{equation}
It is  well known that this regime can also be described within the framework of the WTT and results in a three-wave WKE~\cite{DNPZ,SNbook}. In this work, we will observe such an acoustic regime when the condensate fraction reaches a high value close to one. We will examine the realisability of the Bogoliubov dispersion relation Eq.\eqref{bog}, 
but we will not discuss the spectral evolution governed by the three-wave WKE.

\section{Numerical setup}
\label{sec:numerics}

We perform direct numerical simulations (DNS) of the forced-dissipated GPE by using a  pseudospectral method over triply periodic cubic domain $\mathcal{V}$ with volume $(2\pi)^3$ using $N^3_c=512^3$ collocation points. In this method, we evaluate the linear terms in Fourier space and the nonlinear term in the physical space, which we then transform to Fourier space. To this end, we use the GPE solver of our general purpose code ``VIKSHOBHA''~\cite{VSPRE2018BD}, wherein we use the FFTW library to compute Fourier-transform operations~\cite{fftwsite}. A fourth-order Runge-Kutta scheme is used to evolve these equations in time.

Fixing the values of $\rho$, $c$ and $\alpha$ in the absence of forcing and dissipation, fixes the variables $\xi$ and $g$. However, in presence of forcing and dissipation, the volume-averaged density $\langle \rho(t) \rangle$ varies with time which, in turn, makes $c$  and $\xi$  time dependent. Therefore, in our simulations presented here, we choose an initial state such that $\langle \rho \rangle =\rho^*=1$, $c=c^*=1$. Further, in all our runs we choose the hyperviscosity coefficient $\nu_h =10^{-35}$  and use the boundaries of the forcing range at $k_{f1}=130$ and $k_{f2}=131$.
Forcing amplitude $f_0$ and the parameter 
$\alpha$  used in different simulations are shown in Table~\ref{tab:runs}.  Moreover, we use these to construct  units of: length $\xi^*=\sqrt{2}\alpha/c^*$, time $t^* = \xi^*/c^*$ (not to be confused with the blowup time $t_{\star}$) and energy density $E^* = \rho^*{c^*}^2$. Note that $\xi^*$ is  the initial healing length, which is used as a reference length scale for the run (the true healing length evolves in time).

\begin{table}
	\large
	\begin{center}
		
		\begin{tabular}{l| c| c| c |c} 
			\hline\hline
			{\tt Run} & $\alpha$  & ${\xi^*}$ & $g={c^{*}}^2/2\alpha\rho^*$ & $f_0$ \\
			\hline
			{\tt Run1}  & $0.0347$ & $0.0491$& $14.41$ & $10^8$         \\
			{\tt Run2}  & $0.0347$ & $0.0491$& $14.41$ & $10^7$         \\
			{\tt Run3}  & $0.0087$ & $0.0123$& $57.47$ & $10^8$         \\
		\end{tabular}
	\end{center}
\caption{\small 
The parameter $\alpha$ fixes the quantum of circulation $4\pi\alpha=h/m$ for a particular run. Parameters $\alpha$, $c^*=1$ and $\rho^*=1$ fix the length scale $\xi^*$ ( $\xi$ for the initial state). $f_0$ is the forcing amplitude.
}
\label{tab:runs}
\end{table}

\vspace{.5cm}

\section{Results}
\label{sec:results}

\begin{figure*}
	\resizebox{\textwidth}{!}{%
		\includegraphics[scale=0.4]{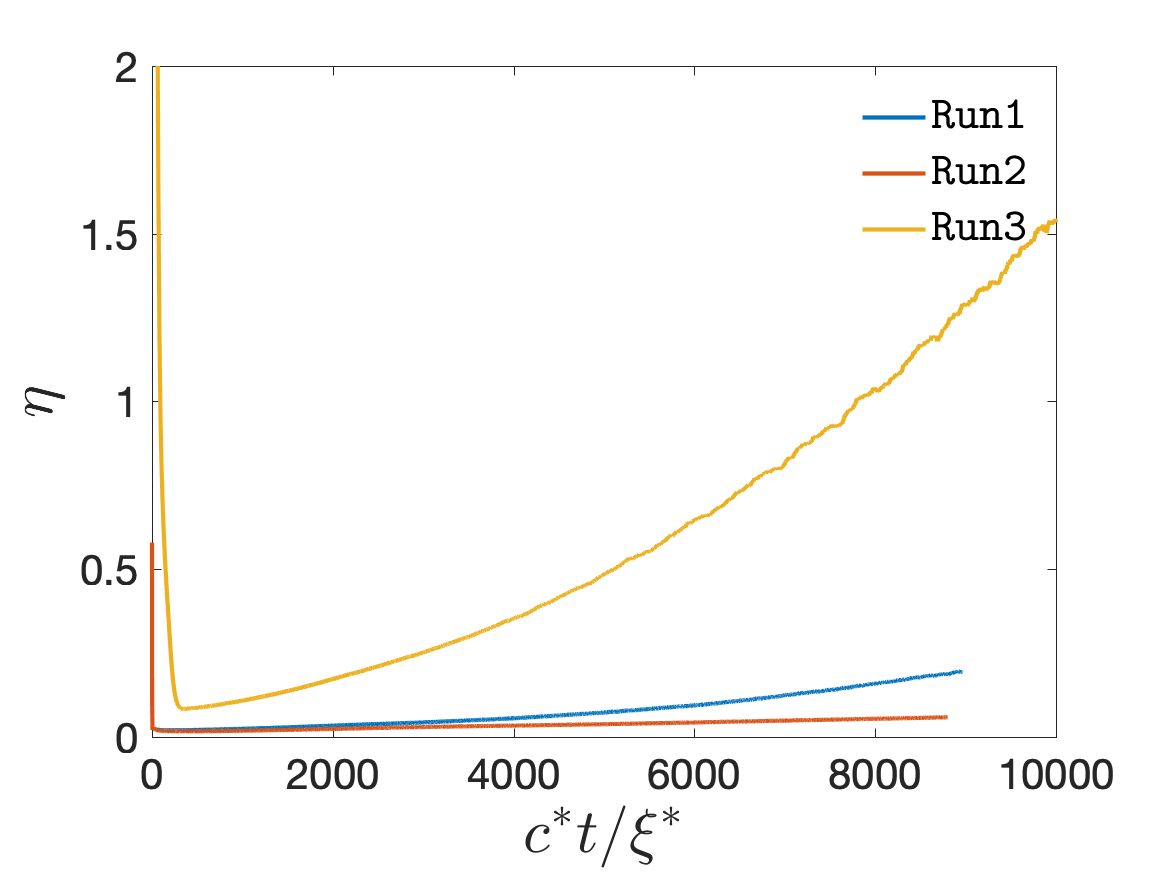}
		\put(-190,130){\bf\large (a)}	
		\includegraphics[scale=0.4]{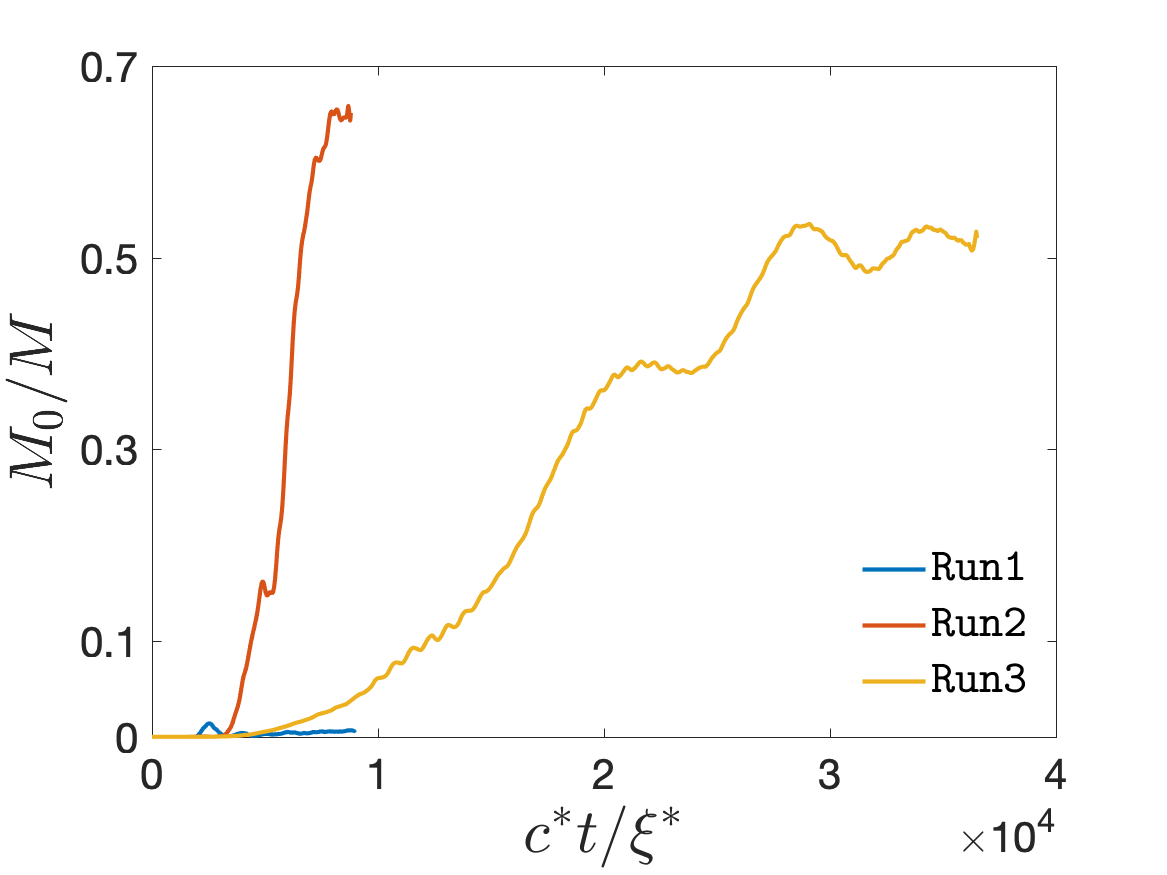}
		\put(-190,130){\bf\large (b)}	
	}
	\caption{Temporal evolution of: (a) nonlinearity parameter $\eta=E_{4}/E_{2}$, (b) the condensate fraction $M_0/M$.}
	\label{Fig:etaM}
\end{figure*}

First of all, we note that larger forcing amplitude $f_0$ and smaller coefficient $\alpha$ (hence larger $g$) mean stronger initial degree of nonlinearity of the system. Thus, among the runs listed in the Table~\ref{tab:runs}, {\tt Run3} has the highest and {\tt Run2} has the lowest nonlinearity. This agrees with Fig.~\ref{Fig:etaM}~(a) where the evolution of the ratio of the quadric to quadratic parts of the energy, $\eta$, is shown for the three runs. (Recall that the quadratic and the quadric energies correspond to the linear and the nonlinear dynamics respectively.) Nonetheless, after an initial adjustment and up until $t/t^* \approx 2000$ even {\tt Run3} remains relatively weakly nonlinear with $\eta <0.2$. Nonlinearities of the runs {\tt Run1} and {\tt Run2} remain rather weak through their entire durations (very weak for the {\tt Run2}). Note that the normalised durations of the runs {\tt Run1} and {\tt Run2} were much less than the one of the {\tt Run3} because their respective values of the characteristic time $t^*$ are much longer.

Figure 1(b) shows the evolution of the condensate fraction, $M_0/M$, in the $k=0$ mode for the runs $\tt Run1-Run3$. We find that in $\tt Run2$ the condensate fraction unexpectedly grows to a large value ($M_0/M \sim 0.65$), even though it has the lowest value of nonlinearity among the three runs. This observation is counter-intuitive, we discuss its plausible explanation later in this section. It must also be mentioned that even though the condensate fraction is high, its absolute value remains rather small ($\rho_0 \sim 0.1$ with total density $\rho \sim 0.2$). Moreover, we will see that it does not affect the dispersion properties of the uncondensed excitations ($k\ne 0$). $\tt Run3$ exhibits second strongest condensate fraction ($M_0/M \sim 0.5$), despite being strongly nonlinear the condensate growth is smaller than what is observed in $\tt Run2$ (when the growth interval is measured in units of $t^*$). However, the condensate fraction in case of $\tt Run1$ (with intermediate nonlinearity) remained vanishingly small throughout the duration of the run.

\begin{figure*}
	\resizebox{\textwidth}{!}{%
		\includegraphics[scale=0.49]{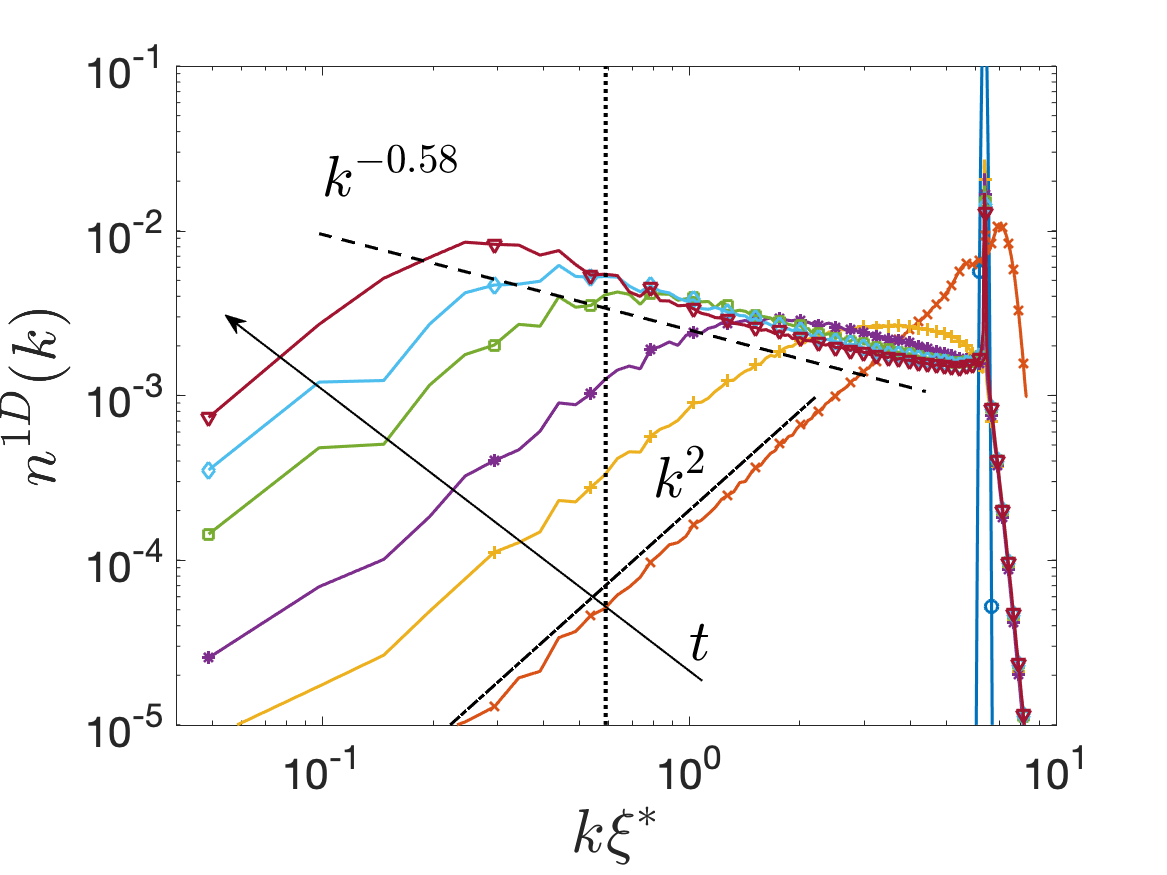}
		\put(-75,50){\bf\large (a)}
		\includegraphics[scale=0.49]{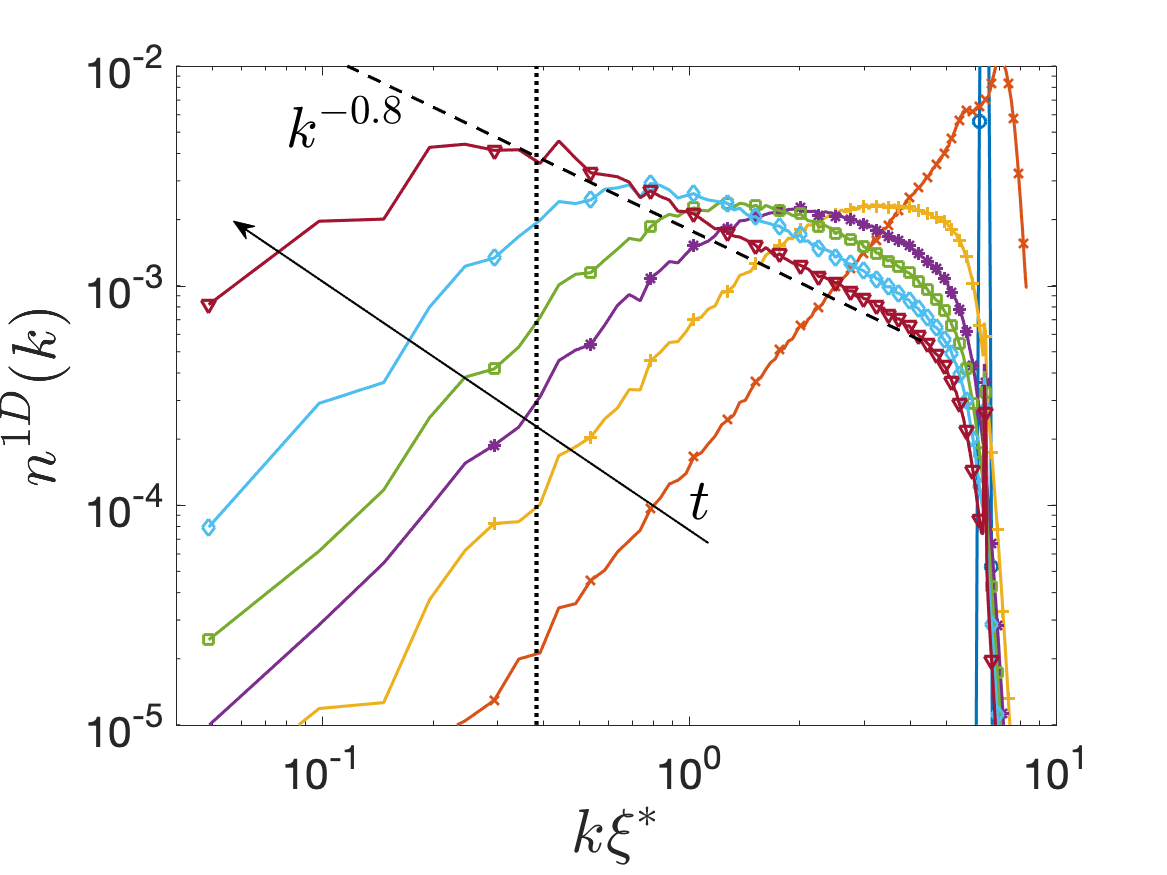}
		\put(-75,50){\bf\large (b)}	
		\includegraphics[scale=0.49]{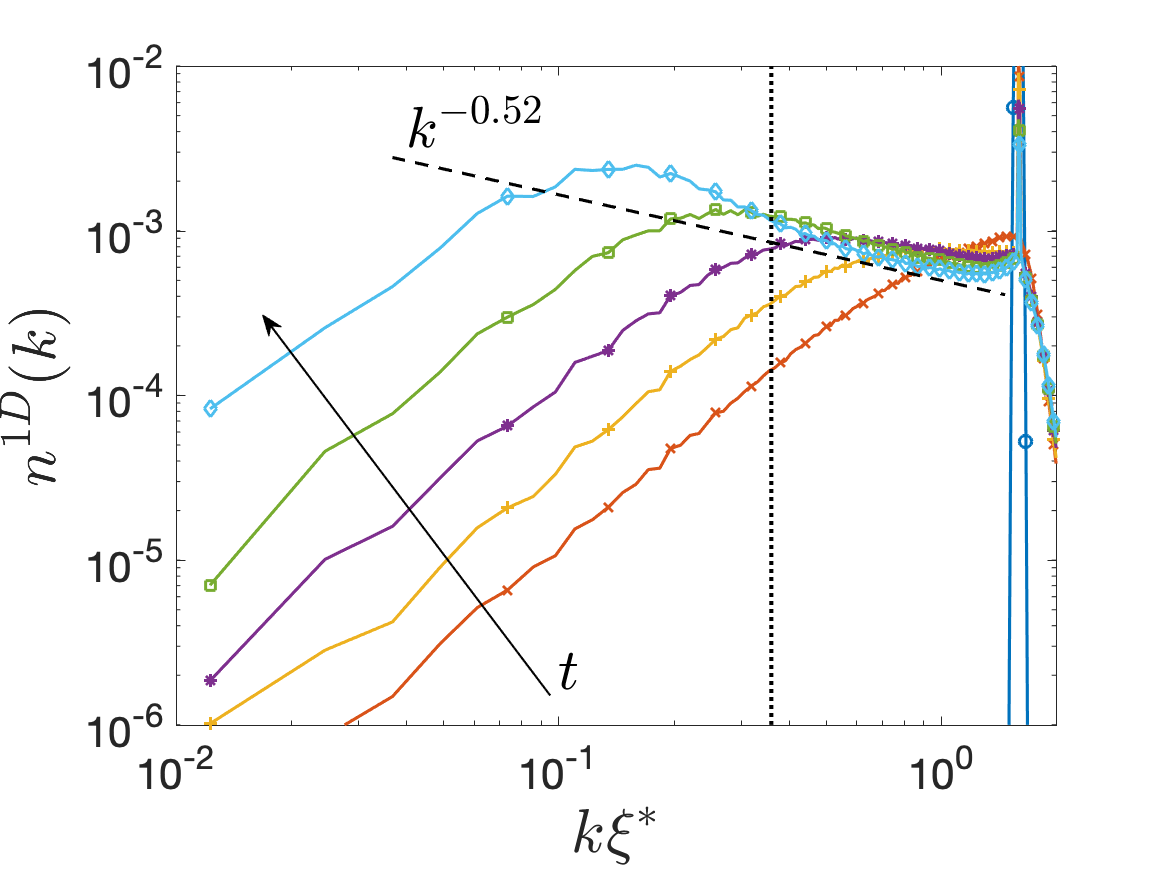}
		\put(-75,50){\bf\large (c)}
	}
	\caption{Temporal evolution of  spectra $n^{1D}(k)$ vs. $k$ at early stages of our numerical simulations. Plots (a), (b) and (c)  correspond to  $\tt Run 1$ ($0 \leq t\,c^*/\xi^* \leq 1.93\times 10^3$), $\tt Run 2$ ($0 \leq t\,c^*/\xi^* \leq 2.54\times 10^3$) and $\tt Run 3$ ($0 \leq t\,c^*/\xi^* \leq 2.0 \times 10^3$), respectively. The vertical dotted line indicates the position of $k\xi(t)\sim 1$ at the latest moment of time. 
		The various slopes shown here  by dashed lines represent power law fits in the interval $
		(1/\xi, k_{f1}).$ 	Dot-dashed line in (a) shows the $k^2$ scaling at small wave numbers.}
	\label{Fig:occpspecearly}
\end{figure*}

\begin{figure*}	
	\resizebox{\textwidth}{!}{%
		\includegraphics[scale=0.49]{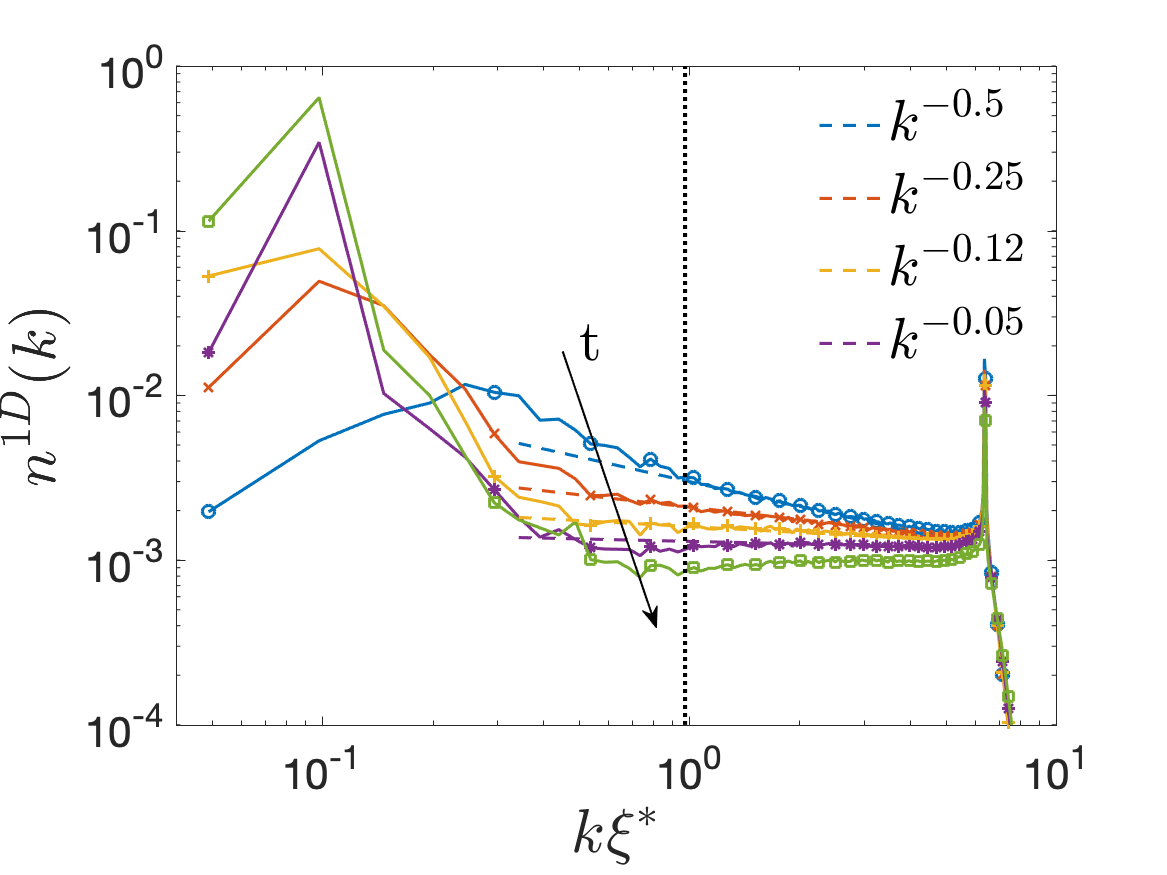}	
		\put(-75,40){\bf\large (a)}
		\includegraphics[scale=0.49]{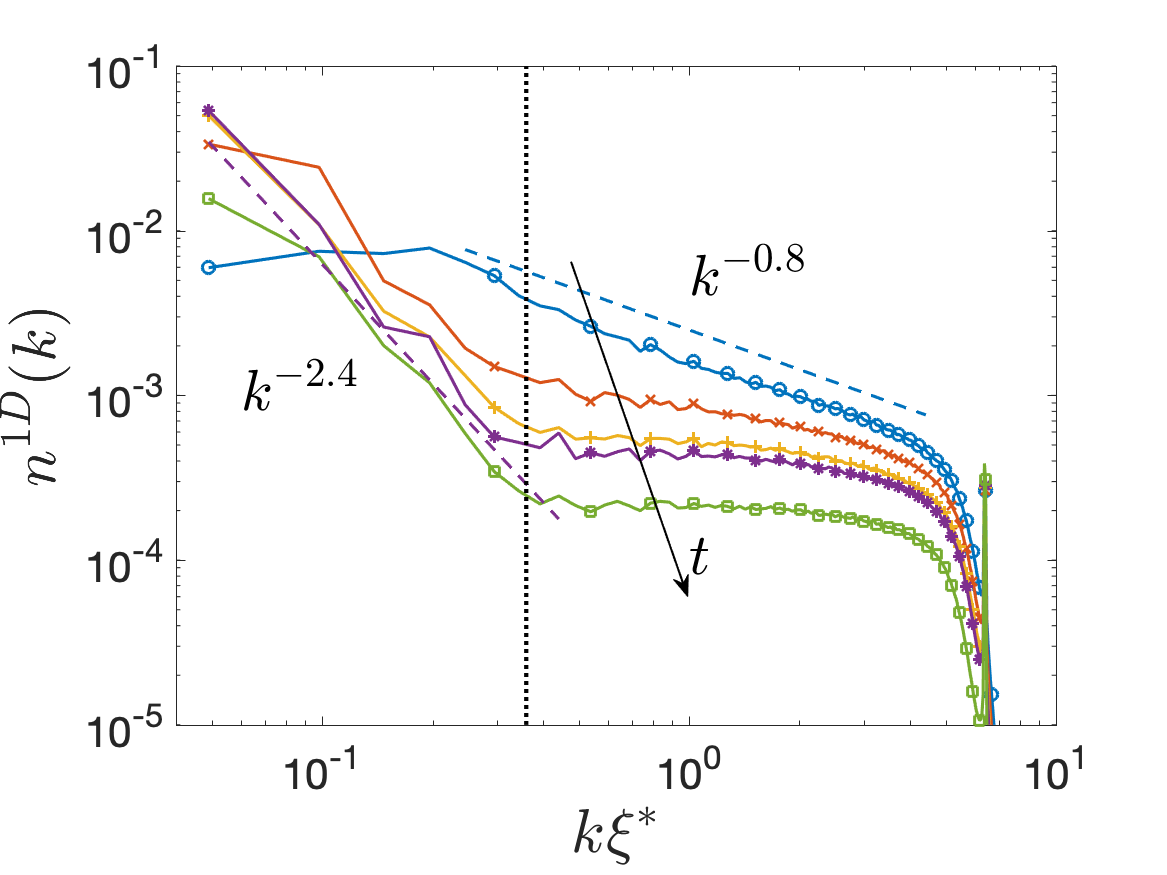}	
		\put(-75,40){\bf\large (b)}
		\includegraphics[scale=0.49]{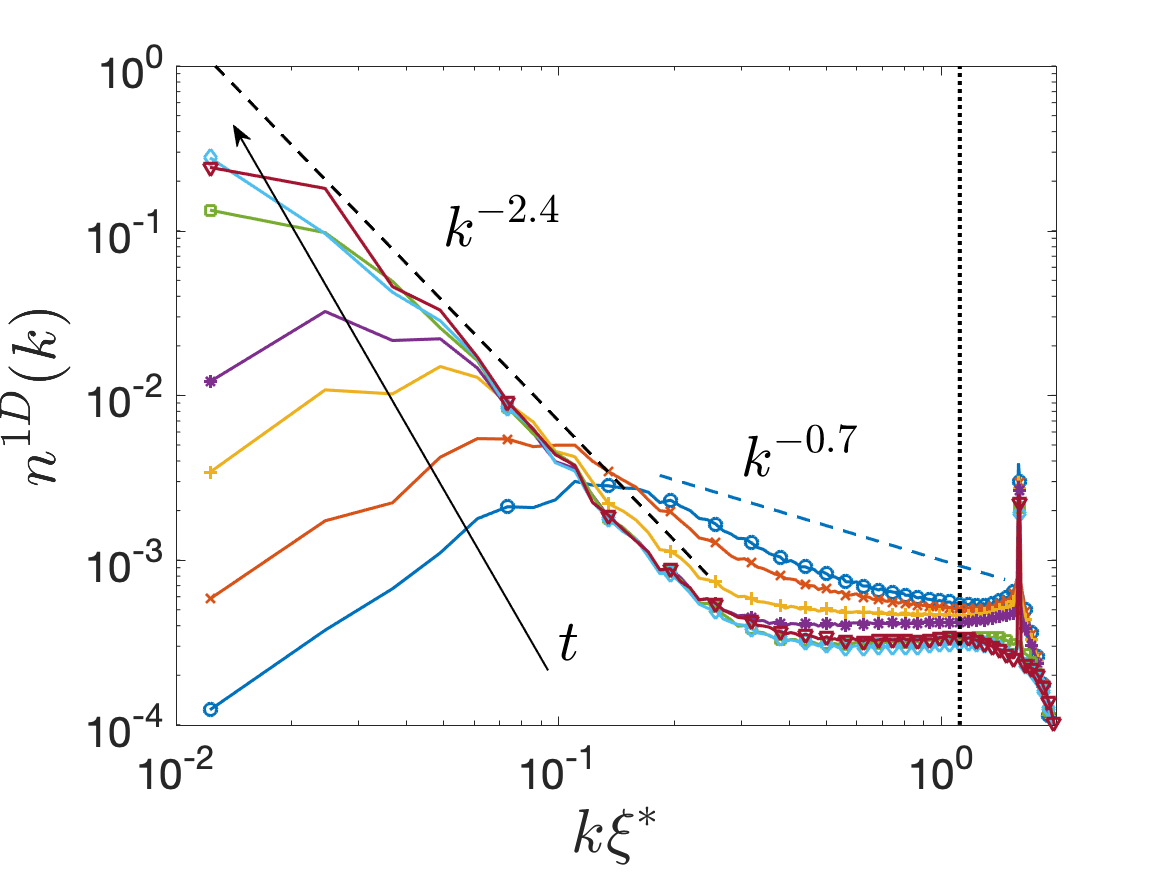}	
		\put(-75,40){\bf\large (c)}
	}
	\caption{Temporal evolution of  spectra $n^{1D}(k)$ at later stages of our numerical simulations. Plots (a), (b) and (c) correspond to  $\tt Run 1$ ($2.03 \times 10^3 \leq t\,c^*/\xi^* \leq 8.95 \times 10^3$), $\tt Run 2$ ($3.05 \times 10^3 \leq t\,c^*/\xi^* \leq 8.80 \times 10^3$) and $\tt Run 3$ ($2.08 \times 10^3 \leq t\,c^*/\xi^* \leq 3.64 \times 10^4$), respectively. The vertical dotted line indicates the position of $k\xi(t)\sim 1$. 
	}	
	\label{Fig:occpspeclate}
\end{figure*}

In Figs.~\ref{Fig:occpspecearly} and \ref{Fig:occpspeclate} we show 1D wave-action spectra at different moments of time computed for the runs $\tt Run1$ - $\tt Run3$. We obtain 1D wave-action spectrum $n^{1D}(k,t)$ by summing the 3D spectra {$\sim|\hat \psi({\bf k}, t)|^2$} in a spherical shell of radius $k$ and thickness $\delta k$ and dividing by $\delta k$. In other words, $n^{1D}(k,t)$ is the wave action density in $k =|{\bf k}|$; the angle dependence is erased, as the system is expected to be approximately statistically isotropic. Moreover, for future reference, we find it pertinent to mention that the power spectrum $n_\omega \sim \omega^{-x}$ written in terms of frequency $\omega$ corresponds to the wave-action spectrum $n^{1D}(k)  \sim k^{2-2x}$.

From the plots in Figs.~\ref{Fig:occpspecearly} and~\ref{Fig:occpspeclate}, we observe that the spectra are in qualitative (and to a varying degree in quantitative) agreement with the WTT results obtained by numerically solving the WKE~(\ref{E1}). Namely, the plots in Fig.~\ref{Fig:occpspecearly} and Fig.~\ref{Fig:occpspeclate} are consistent with the pre-blowup and the post-blowup behaviour. Indeed, the early evolution in Fig.~\ref{Fig:occpspecearly} exhibits emergence of a power law scaling  $n^{1D}(k)  \sim k^{2}$ which corresponds to
$n_\omega  \sim $~const; this is in agreement with the low-frequency asymptotics of the self-similar solution of the WKE. This is followed by the development of a second power law scaling region at the higher-$k$ side of the inverse cascade range with slopes $\sim -0.58$ in {\tt Run1}, $\sim -0.80$ in {\tt Run2} and $\sim -0.52$ in {\tt Run3} that correspond to exponents $x\approx 1.29, 1.40$ and $1.26$, respectively, for $n_\omega$. Note that these slopes are computed over the range $1/\xi < k < k_{f1}$ because for $k \lesssim 1/\xi$ the nonlinearity fails to be weak. It is evident that these exponents are close to the values obtained for the self-similar solution of the WKE, $x_{\star}\approx 1.22-1.24$, for details see Sec.~\ref{sec:WTP}. Later, we will discuss the possible reason why {\tt Run3} gives the closest result.

The late-time behaviour shown in Fig.~\ref{Fig:occpspeclate} is consistent with the post-$t_{\star}$ behaviour of the WKE evolution. For all the three runs, at the higher-$k$ end of the inverse cascade range we observe formation of a plateau $n^{1D}(k) \sim \text{const.}$ which corresponds to the thermodynamic energy equipartition $n_\omega \sim 1/\omega$. At the lowest wave numbers we observe a sharply peaked spectrum that appears to decay as $k^{-2.5}$ instead of the predicted infinitely thin (Dirac delta) distribution at  $k=0$ mode representing the condensate. It is natural that such an idealized distribution is not observed in our DNS because the low-k dynamics is affected by both the finite size of the system and an enhanced nonlinearity. Thus, the condensate emerging in our simulations is imperfect: it contains space-time dependent strongly-nonlinear coherent structures, we will revisit this again later.
The slope $-2.5$ for the 1D waveaction spectrum is rather close to the low-k slope $-3$ ($-5$ for the 3D waveaction spectrum) observed in a closed system for a freely decaying initial condition (no forcing or dissipation) in~\cite{nowak2012nonthermal}, where a scaling argument suggested that the  $-3$ slope should be observed for strong turbulence.

 Also note that both the condensate and the plateau region in spectrum at late times for {\tt Run3} are located at the wave numbers well below $1/\xi$. This means that the influence of the condensate onto the high-$k$ modes is significant: these modes behave as sound waves and, therefore, are not described by the four-wave WKE~(\ref{E1}); see Sec.~\ref{sec:WTP} . 
However, it is interesting to observe that the energy equipartition spectrum in this case as well is given by  $n^{1D}(k) \sim \text{const}$, which for the present state is called the Bogoliubov spectrum~\cite{BogoliubovJrBook,DNOQuantumcascade,vmrnjp13,SNBKT}.

\begin{figure*}
	\includegraphics[scale=0.30]{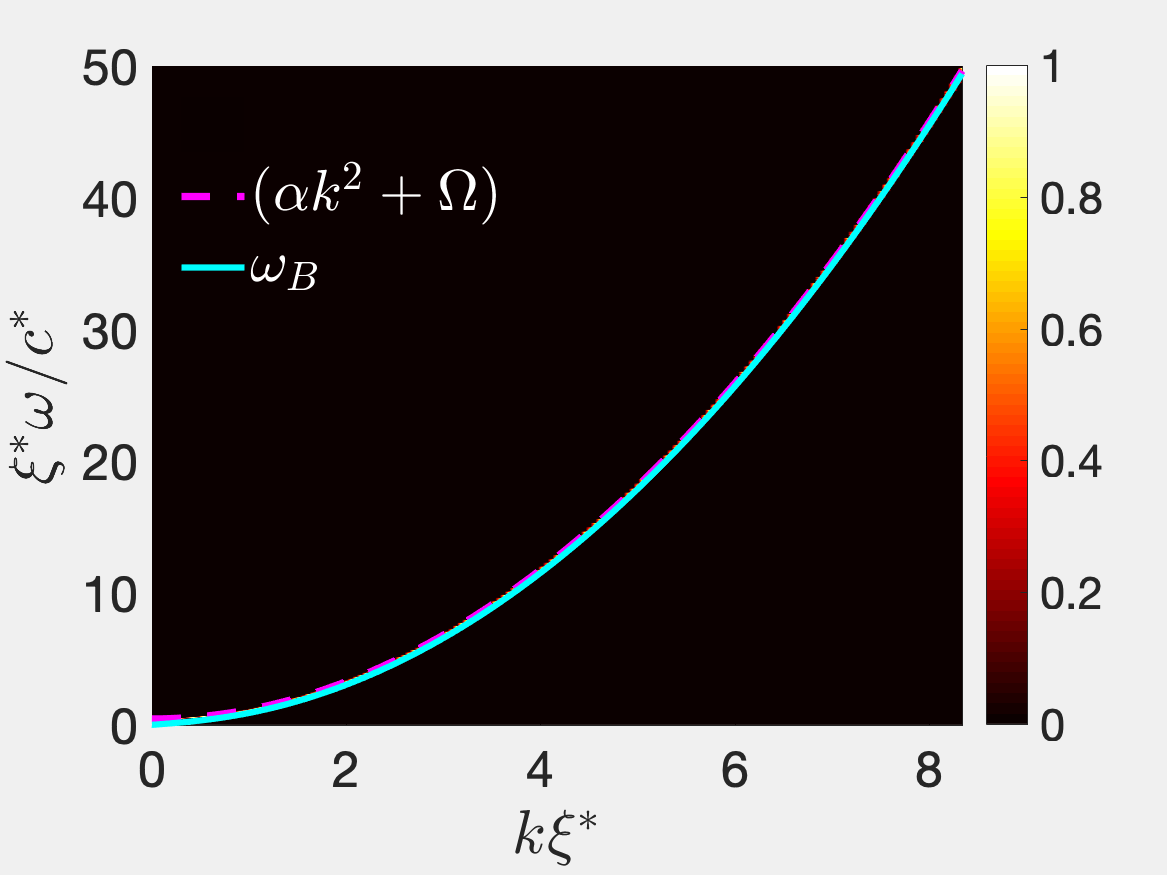}
	\put(-55,25){\textcolor{white}{\bf\large (a)}}	
	\includegraphics[scale=0.30]{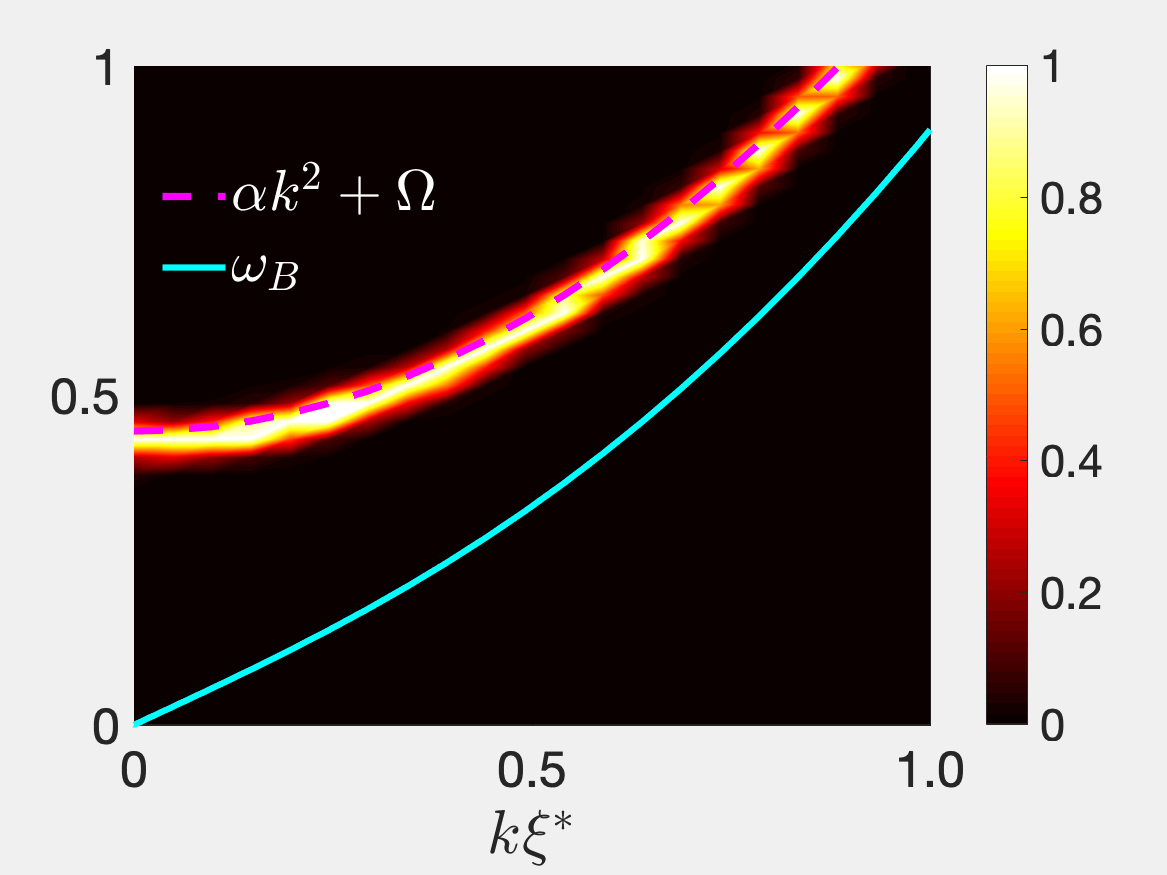}	
	\put(-60,25){\textcolor{white}{\bf\large (a1)}}		
	\includegraphics[scale=0.30]{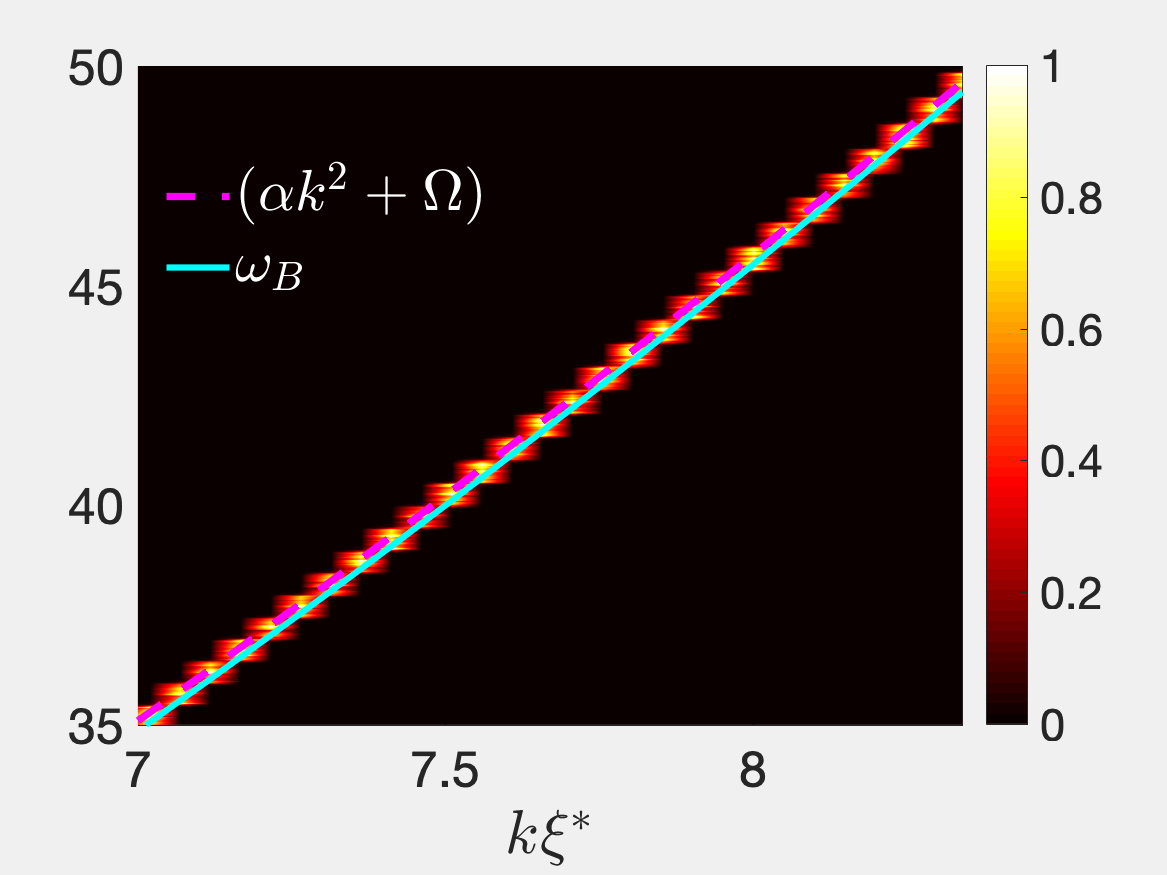}
	\put(-60,25){\textcolor{white}{\bf\large (a2)}}	
	\caption{Plots of spatio-temporal spectra for the $\tt Run1$ computed over the time interval $1.14\times 10^3 \leq Tc^*/\xi^* \leq 1.37\times 10^3$. (a) The full spectrum; (a1) and (a2) zoom at small and large wave numbers, respectively. The forcing  range is  $k_{f}\xi^* \in (6.3830, 6.4321)$.  During the above time interval $n^{1D}(k) \sim k^{-0.58}$, see Fig.~\ref{Fig:occpspecearly} (a).}
	\label{ST1}
\end{figure*}

\begin{figure*}	
	\includegraphics[scale=0.30]{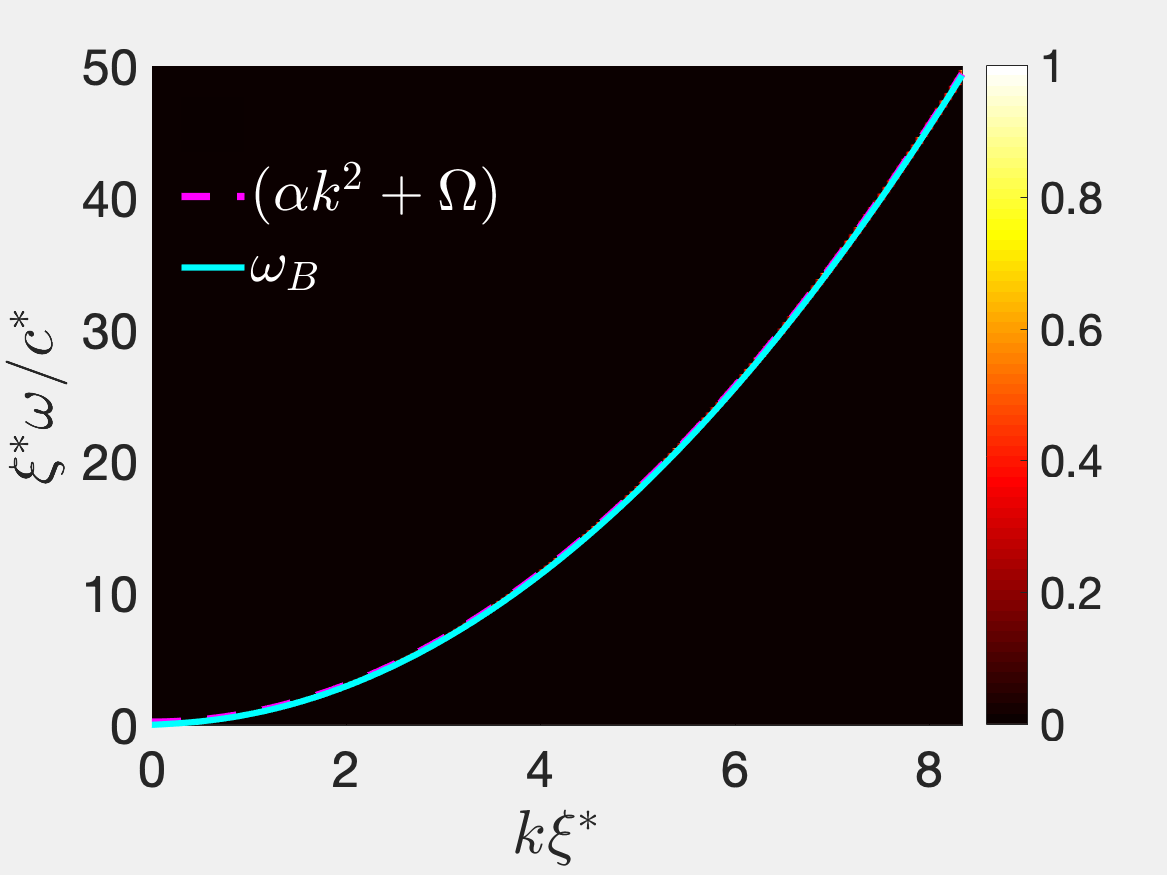}
	\put(-55,25){\textcolor{white}{\bf\large (a)}}	
	\includegraphics[scale=0.30]{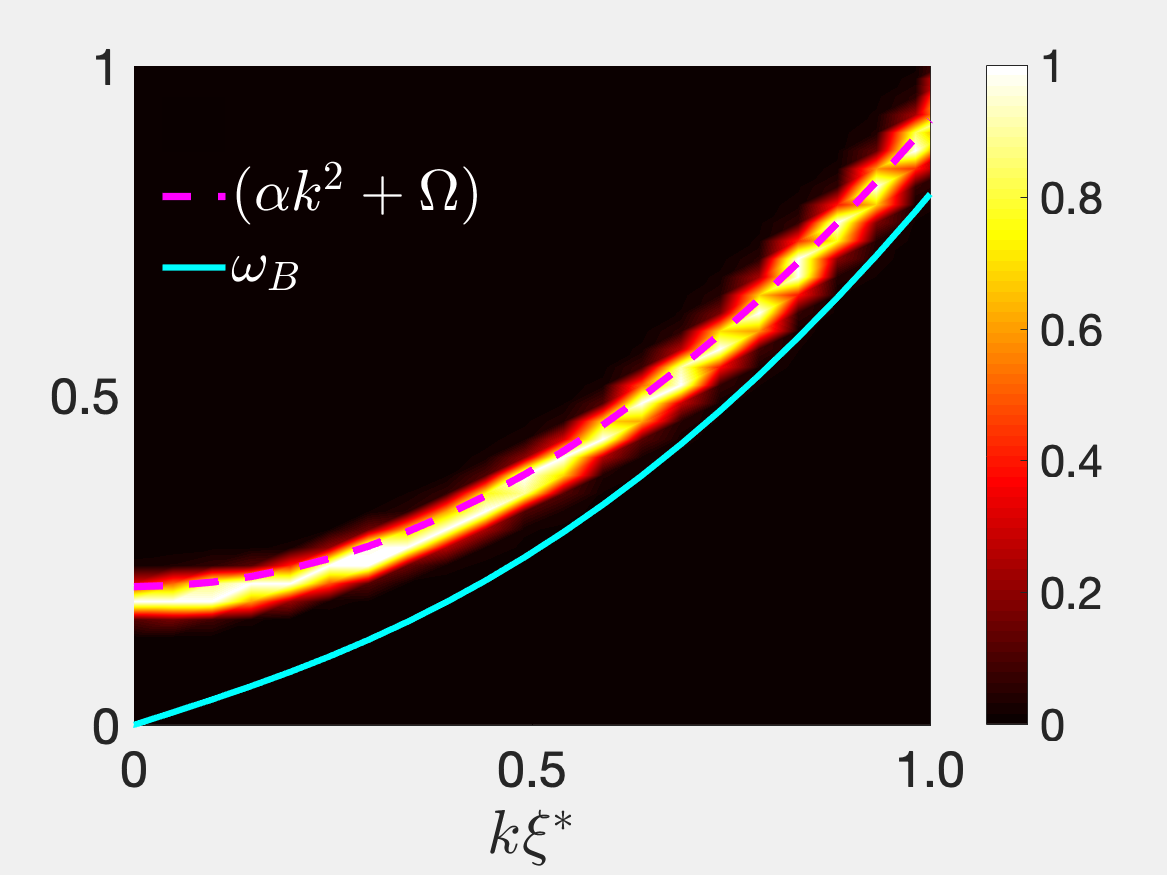}	
	\put(-60,25){\textcolor{white}{\bf\large (a1)}}	
	\includegraphics[scale=0.30]{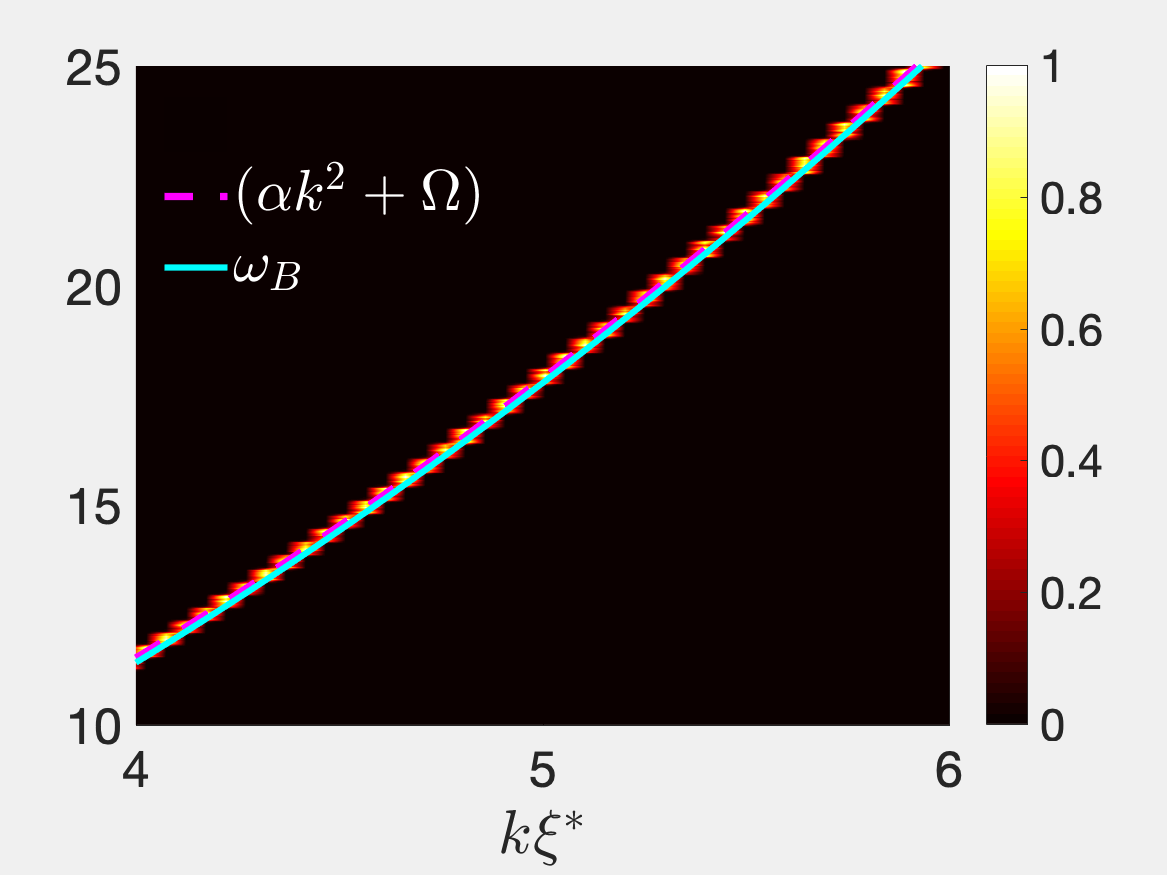}
	\put(-60,25){\textcolor{white}{\bf\large (a2)}}	
	\\
	\includegraphics[scale=0.30]{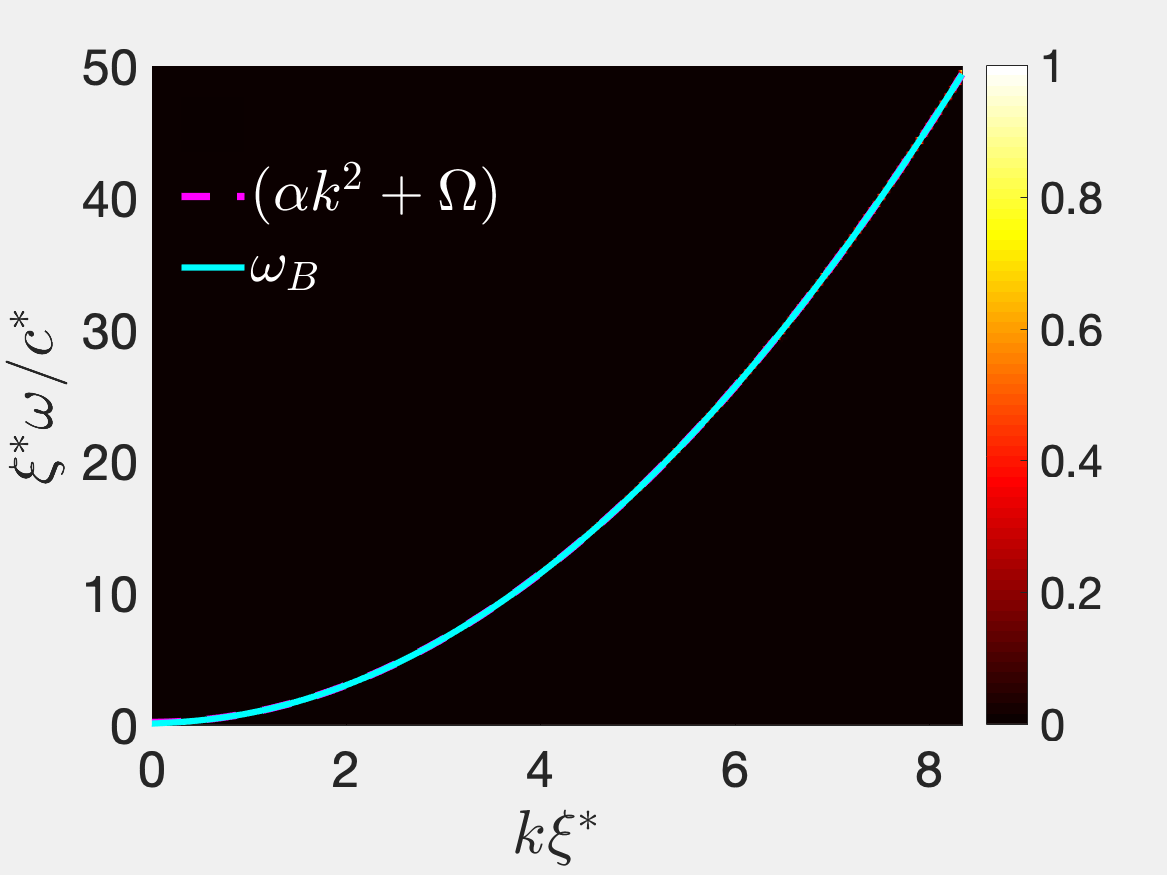}
	\put(-55,25){\textcolor{white}{\bf\large (b)}}	
	\includegraphics[scale=0.30]{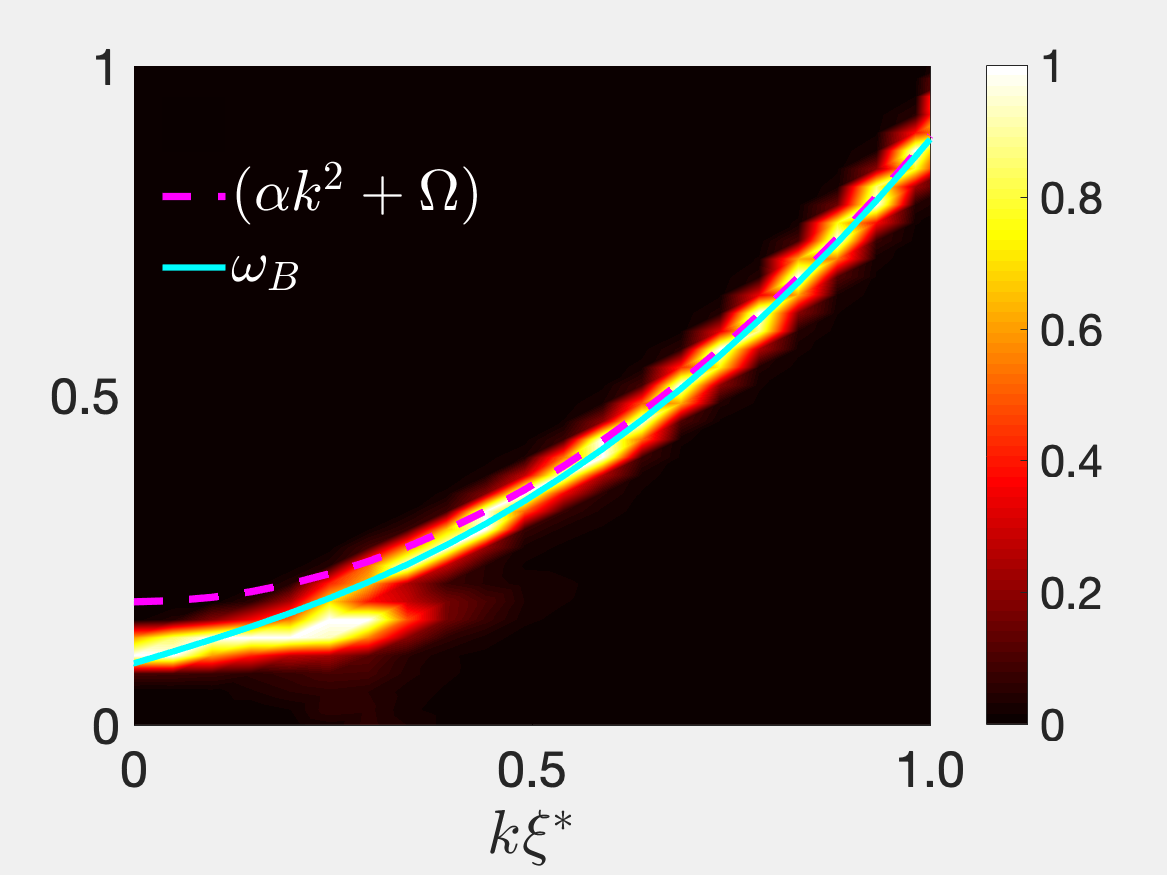}	
	\put(-60,25){\textcolor{white}{\bf\large (b1)}}	
	\includegraphics[scale=0.30]{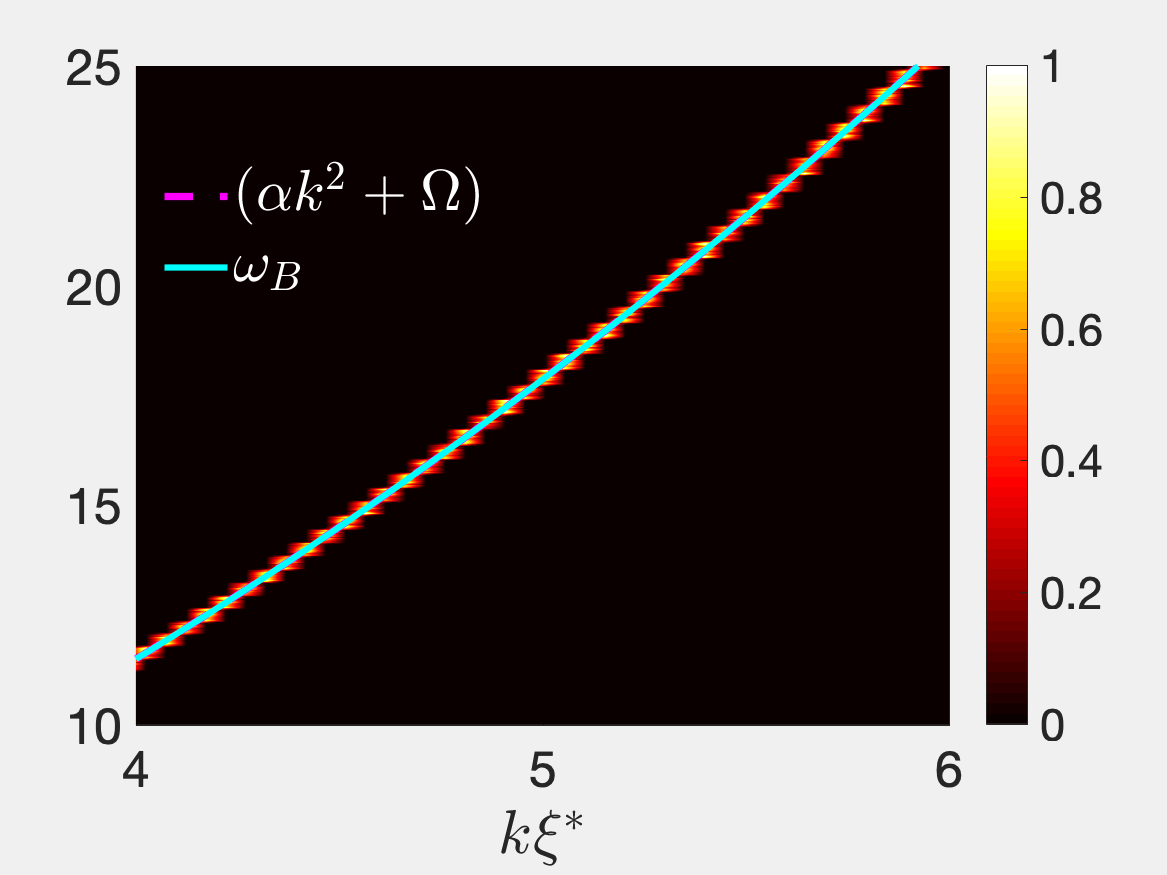}	
	\put(-60,25){\textcolor{white}{\bf\large (b2)}}	
	\caption{Plots of spatio-temporal spectra for the $\tt Run2$. (a) ST spectra computed over the time interval $2.45\times10^3 \leq Tc^*/\xi^* \leq 2.67 \times 10^3$. (a1) and (a2) show a zoom at small and large wave numbers, respectively. During the above time interval $n^{1D}(k) \sim k^{-0.8}$, see Fig.~\ref{Fig:occpspecearly} (b). (b) ST spectra computed over the time interval $6.52 \times 10^3 \leq Tc^*/\xi^* \leq 6.75 \times 10^3$; (b1) and (b2) show a zoom at small and large wave numbers, respectively. The forcing  range is  $k_{f}\xi^* \in (6.3830, 6.4321)$.}
	\label{ST2}
\end{figure*}

\begin{figure*}	
	\includegraphics[scale=0.30]{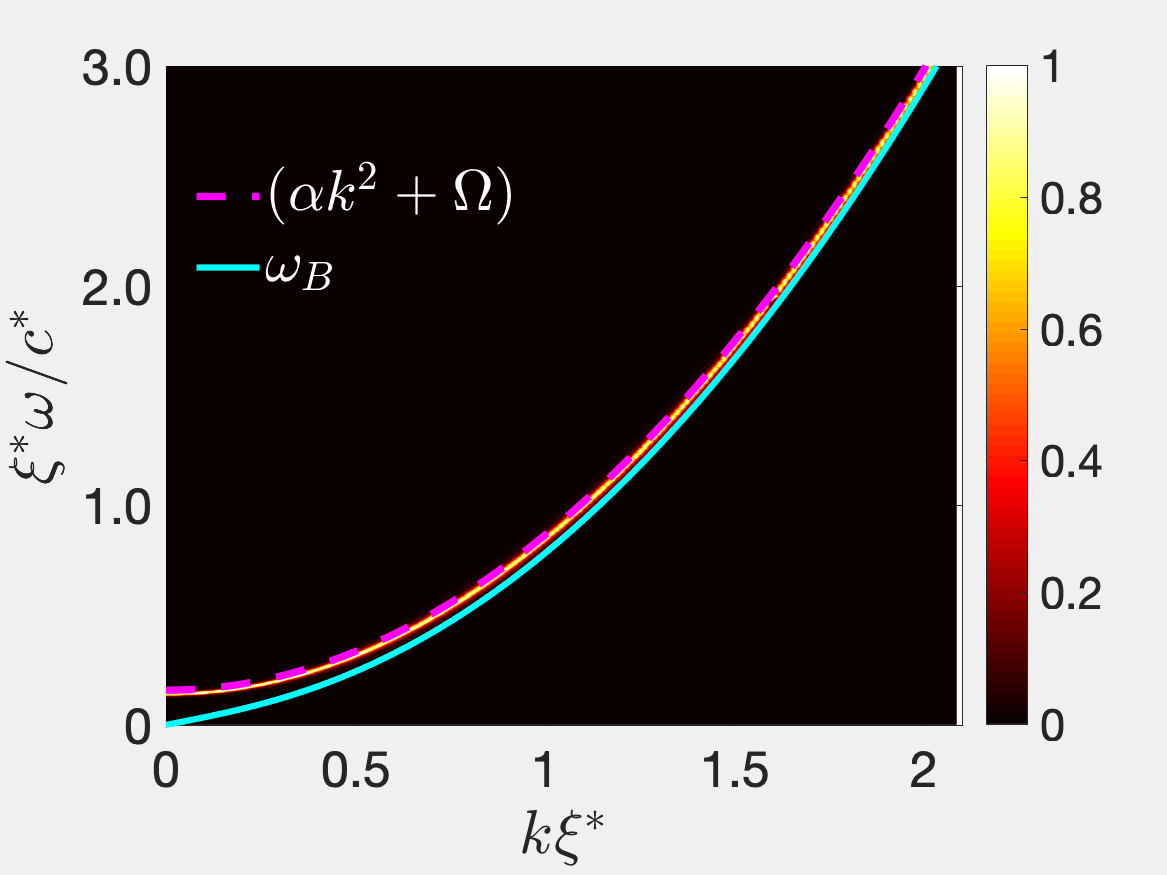}
	\put(-55,25){\textcolor{white}{\bf\large (a)}}	
	\includegraphics[scale=0.30]{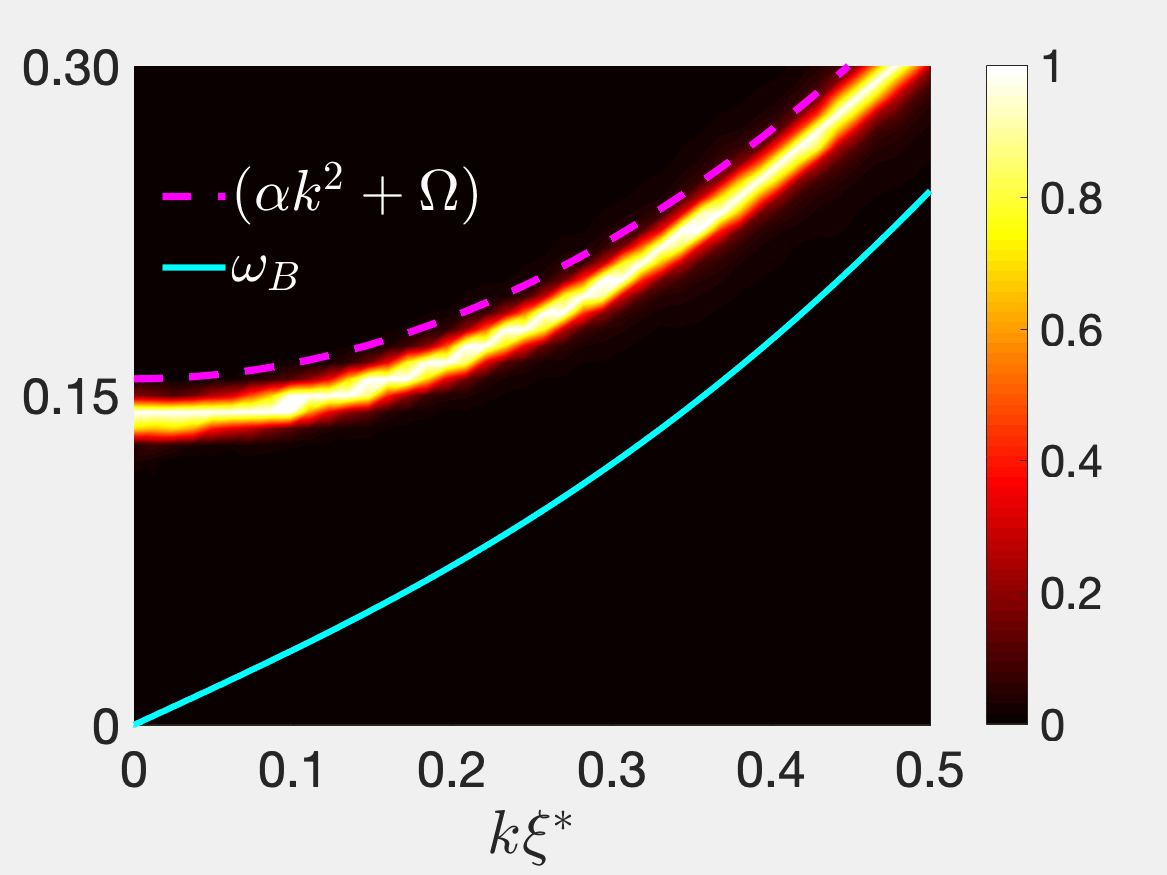}
	\put(-60,25){\textcolor{white}{\bf\large (a1)}}		
	\includegraphics[scale=0.30]{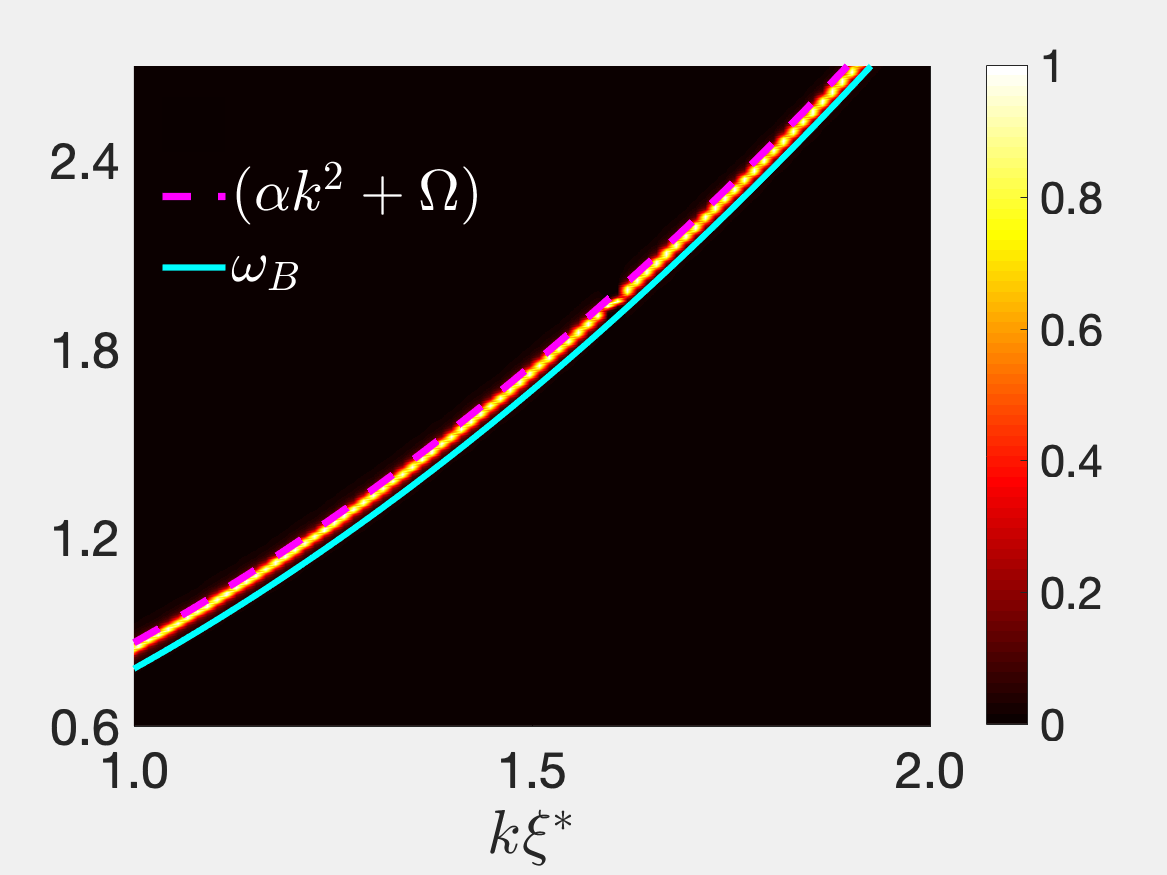}
	\put(-60,25){\textcolor{white}{\bf\large (a2)}}	
	\\
	\includegraphics[scale=0.30]{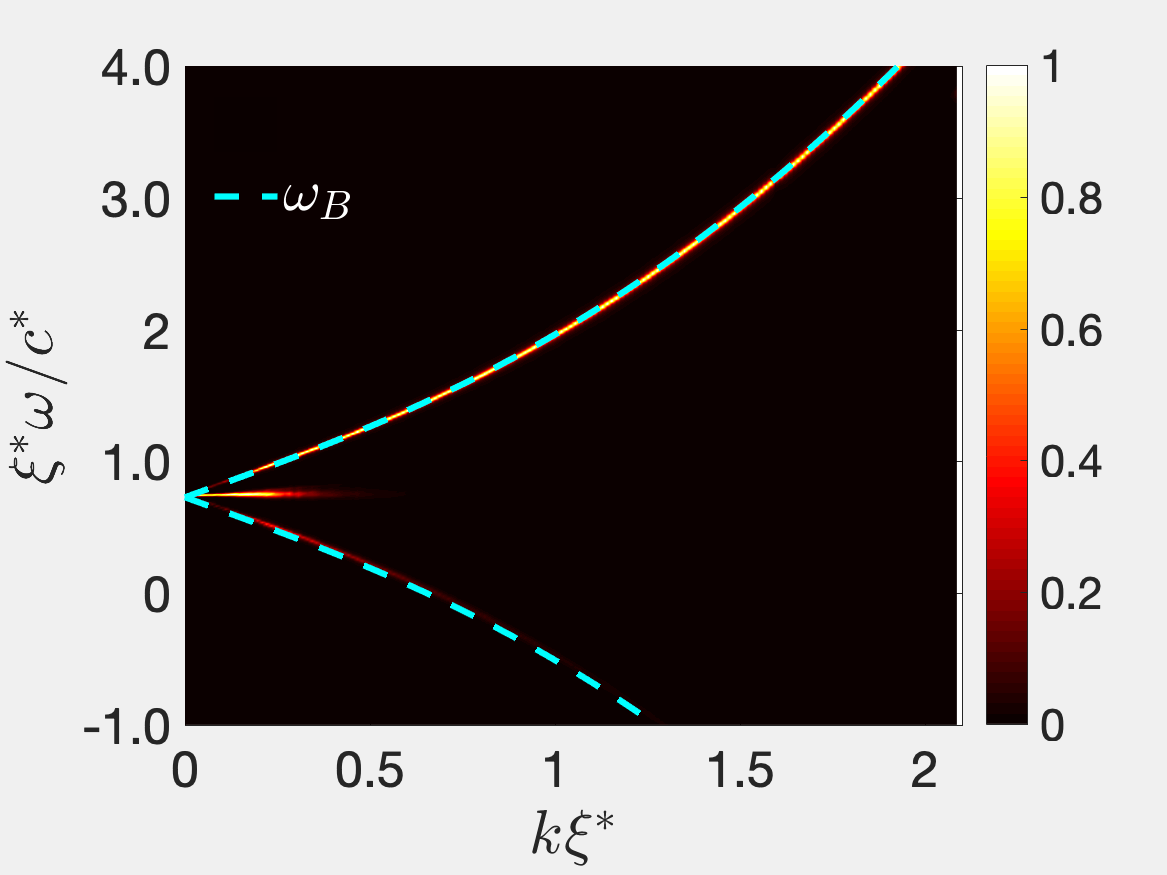}
	\put(-55,25){\textcolor{white}{\bf\large (b)}}	
	\includegraphics[scale=0.30]{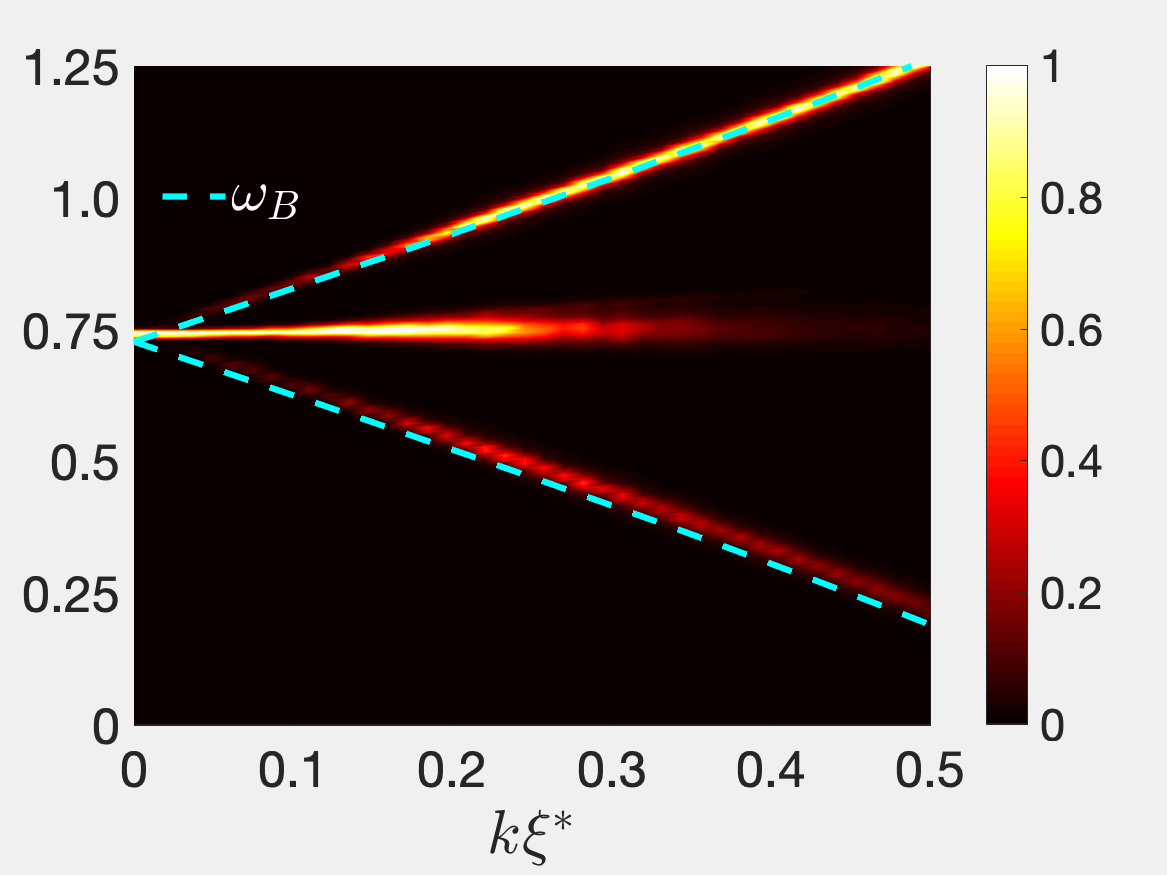}	
	\put(-60,25){\textcolor{white}{\bf\large (b1)}}	
	\includegraphics[scale=0.30]{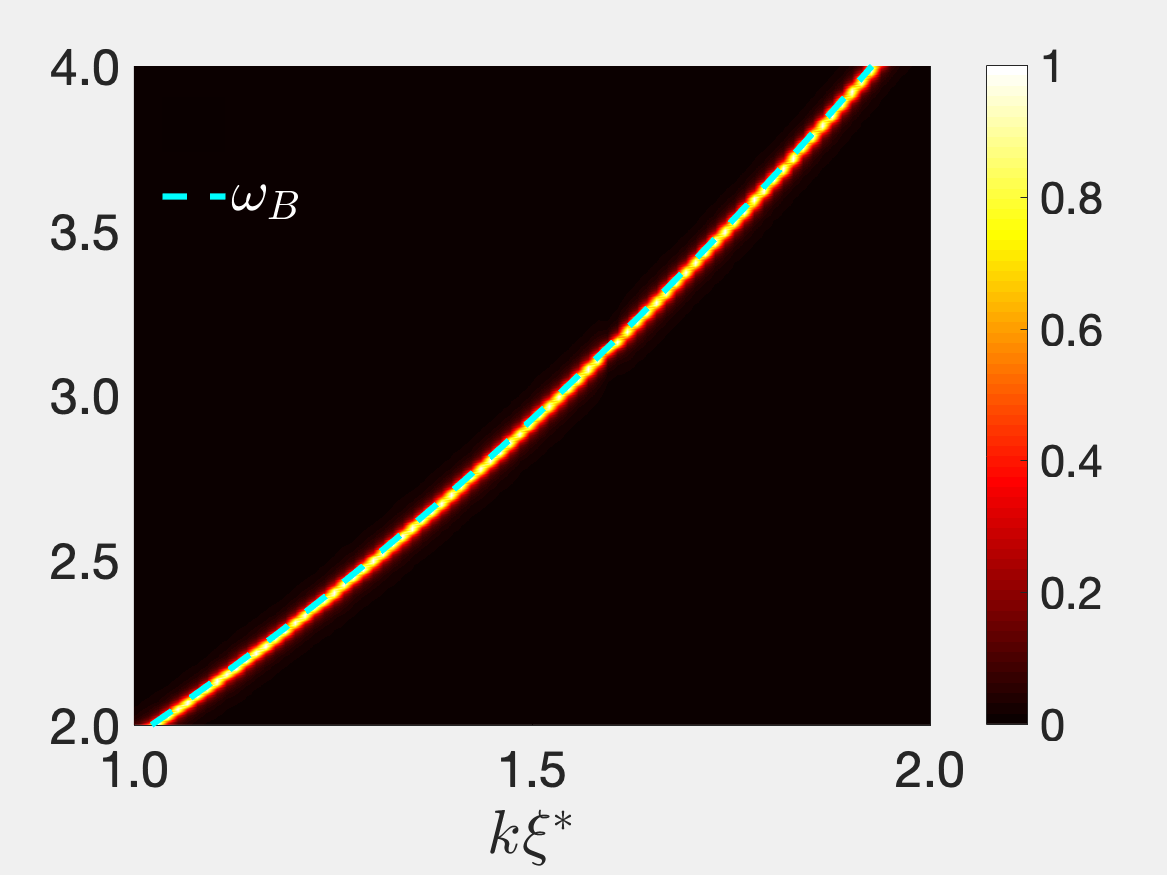}	
	\put(-60,25){\textcolor{white}{\bf\large (b2)}}	
	\caption{Plots of spatio-temporal spectra for the $\tt Run3$. (a) ST spectra computed over the time interval $1.30 \times 10^3 \leq Tc^*/\xi^* \leq 2.14 \times 10^3$; (a1) and (a2) show a zoom at small and large wave numbers respectively. During the above time interval $n^{1D}(k) \sim k^{-0.52}$, see Fig.~\ref{Fig:occpspecearly} (c). (b) ST spectra computed over the time interval $2.80 \times 10^4 \leq Tc^*/\xi^* \leq 2.89 \times 10^4$; (b1) and (b2) show a zoom at small and large wave numbers, respectively. The forcing  range is  $k_{f}\xi^* \in (1.5990, 1.6113)$. 
	}
	\label{ST3}
\end{figure*}

Next we make use of the spatio-temporal (ST) spectra of field $\psi(\r,t)$ to characterize and emphasize the differences between different stages of the evolution of our three runs $\tt Run1 - Run3$. ST spectrum gives the power spectrum of a given quantity as a function of wave number and frequency. In Figs.~\ref{ST1}--\ref{ST3} we show these ST spectra for the runs $\tt Run1 - Run3$. We compute ST spectra by performing Fourier transform over a finite window of size $T$ which is greater than the linear wave period and less than the characteristic time over which the spectrum evolves. Since the spectrum is nearly isotropic, we choose $k_y=0$ and  perform the Fourier transform in time for each $(k_x,k_z)$; this is followed by angle averaging in the $(k_x,k_z)$-plane to finally obtain ST spectra as a function of $k=\sqrt{k_x^2+k_y^2}$ and $\omega$. Moreover,  plots in Figs.~\ref{ST1}--\ref{ST3} are presented in such a way that at each $k$ we normalise the integral over $\omega$ to the same constant; this allows us to see the details of the spectrum at each $k$ clearly, even though its value varies over many orders of magnitude in the $k$ range.

Before proceeding further, we remark that a weak nonlinearity manifests itself in the form of a narrow distribution of the ST spectrum around the linear dispersion relation $\omega_k=\alpha k^2$, along  with a small nonlinear correction $\Omega=2gM/L^3$. This is indeed what we find during the initial evolution as shown in the plots of Figs.~\ref{ST1} - \ref{ST3}. Late evolution in {\tt Run1} is characterized by very similar plots of ST spectra (omitted here) because the condensate faction remains negligible throughout its duration. On the other hand, late evolution in runs {\tt Run2} and {\tt Run3} is characterized by a large condensate fraction. Hence, it is not surprising that the ST spectra in these cases follow more closely the Bogoliubov dispersion relation Eq.~\eqref{bog} rather than $\omega(k) = \alpha k^2+\Omega$, see the plots (b) and (b1) in Figs.~\ref{ST2}--\ref{ST3}. Furthermore, in these plots  the condensate component shows up as a short piece of horizontal line at small $k$'s and $\omega=g \rho$. The fact that the condensate has short but finite range in $k$ means that it is not perfectly uniform in the physical space, whereas the fact that its ST spectrum is nearly horizontal means that the condensate represents a coherent component which oscillates at approximately the same frequency as a whole.

\begin{figure*}	
	\resizebox{\textwidth}{!}{%
		\includegraphics[scale=0.49]{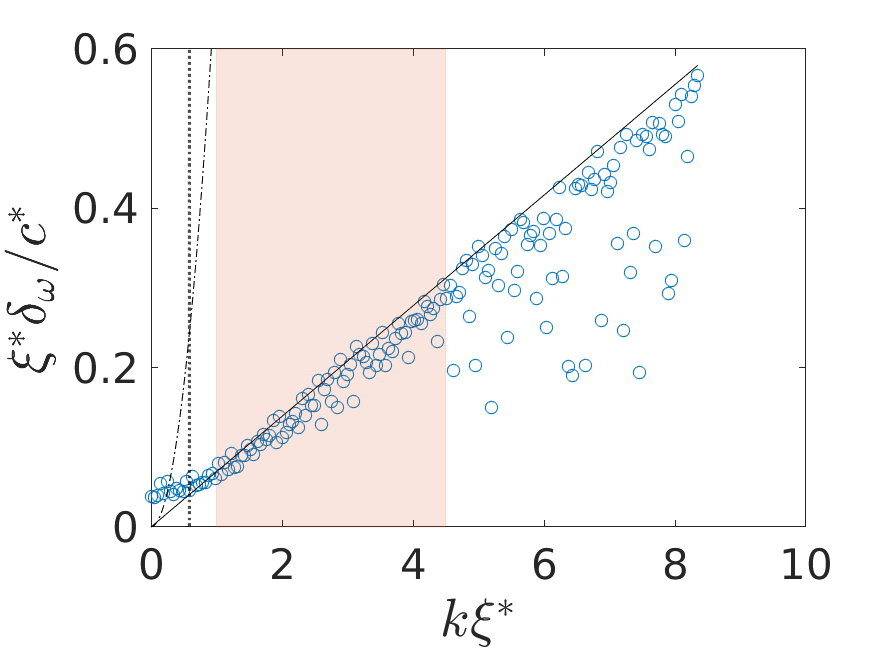}
		\put(-45,40){\bf\large (a)}
		\includegraphics[scale=0.49]{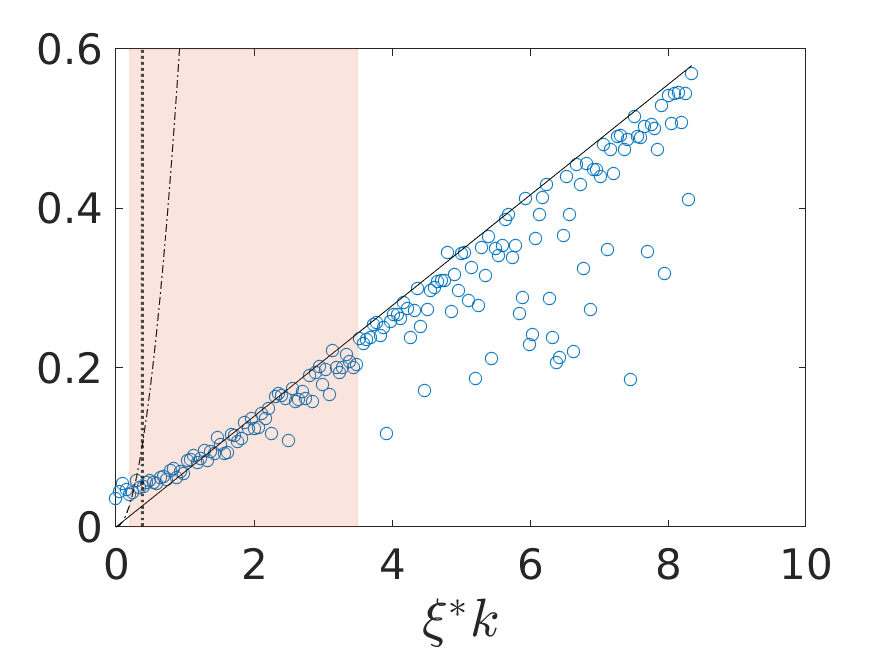}	
		\put(-45,40){\bf\large (b)}
		\includegraphics[scale=0.49]{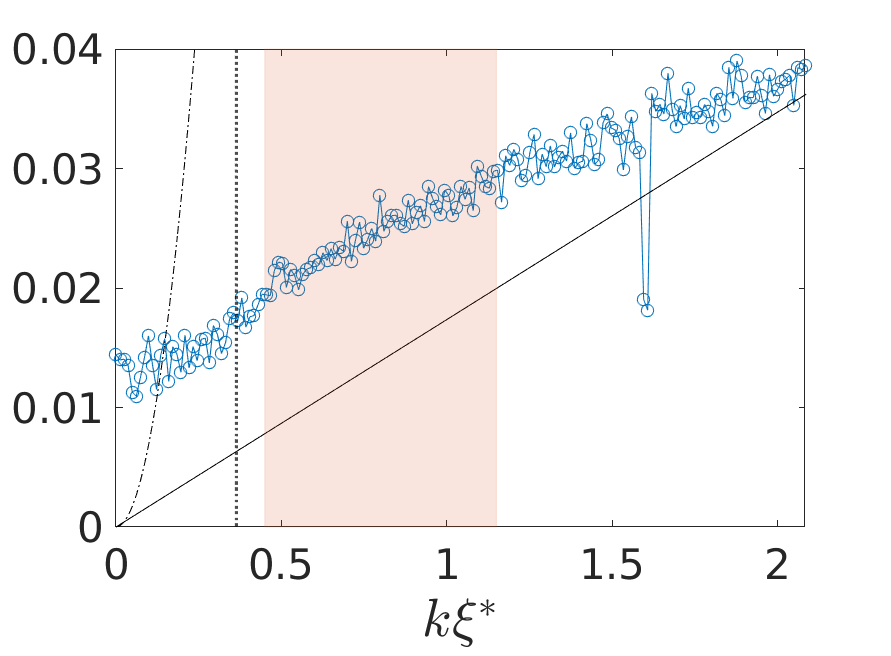}	
		\put(-45,40){\bf\large (c)}	
	}
	\caption{Spectral broadening $\delta_\omega$ vs. $k$ obtained from the spatio-temporal spectra for Runs: (a) $\tt Run1$  $(1.14\times 10^3 \leq tc^*/\xi^* \leq 1.37 \times 10^3)$; (b) $\tt Run2$ $(2.45\times 10^3 \leq tc^*/\xi^* \leq 2.67 \times 10^3)$; and (c) $\tt Run3$ $(1.30\times 10^3 \leq tc^*/\xi^* \leq 2.14 \times 10^3)$.
		The vertical dotted line indicates the position of $k\xi(t)\sim 1$ at time: (a) $tc^*/\xi^* = 1.3\times 10^3$; (b) $tc^*/\xi^* = 2.54\times 10^3$; and (c) $tc^*/\xi^* = 2.0\times 10^3$. Shaded area indicates the approximate region over which power-law scaling are observed in Fig.~\ref{Fig:occpspecearly}.
	}
	\label{broadening}
\end{figure*}

The ST spectra can also be used to measure the nonlinear frequency broadening $\delta_\omega $ and, therefore, directly examine if the conditions of applicability of the WTT are satisfied. We recall from the previous discussion that for the WTT applicability $\delta_\omega $ must satisfy
$2 \alpha k (2\pi/L) = \Delta \omega \ll \delta_\omega \ll \omega_k=\alpha k^2 $. On the ST spectra plots, we define $\delta_\omega $ at each fixed $k$ as the full width at half maximum of the $\omega$-peak. 
Figure~\ref{broadening} (a)-(c) shows the plots of $\delta_\omega $, together with $\omega_k$ and $\Delta \omega$, for the three runs. The time intervals for these plots are chosen to be around the time at which the pre-$t_{\star}$ power-law scalings are present, and the shaded regions indicate the $k$-range over which these scalings were observed. We clearly see that the degree of nonlinearity in these scaling intervals is low, $\delta_\omega < \omega_k$, as required by the WTT. On the other hand, we see that the condition $\delta_\omega > \Delta \omega$ is well satisfied for the {\tt Run3} only. Thus, the applicability of the WTT description is best satisfied in the {\tt Run3}, which explains why this run gives the best agreement with the WKE result for the pre-$t_{\star}$ power law exponent $x_{\star}$ ($1.26$ in the {\tt Run3}  vs $1.22-1.24$ from the WKE simulations). We also recall that the nonlinearity of the {\tt Run3} is the highest among the three runs, and it is the higher nonlinearity that allows it to overcome the effect of discreteness of the $k$-space.  {\tt Run1} and {\tt Run2} are both affected by the $k$-space discreteness. However, the nonlinearity of {\tt Run1} is higher than the one of {\tt Run2}, and it is therefore natural that the pre-$t_{\star}$ power law exponent $x_{\star}$ in {\tt Run1} is closer to the WKE result ($1.28$ and $1.40$ in the {\tt Run1} and {\tt Run2}, respectively,  vs $1.22-1.24$ from the WKE simulations).

It is interesting to note that the nonlinear frequency broadening has qualitatively the same behaviour in the plots (a) and (b) of Figs.~\ref{broadening}, even though the degree of nonlinearity of runs {\tt Run1} and {\tt Run2} are very different: it is in approximate balance with the linear wave frequency spacing, $\delta_\omega \sim \Delta \omega$, which is particularly clear in the scaling range. A possible explanation for this behavior is a sandpile effect which was initially discussed in the context of water waves~\cite{SN2006sandpile}. In forced systems, the turbulent cascade (the inverse cascade of mass in our case) is initially impeded by the discreteness of the Fourier space because there are less frequency and wave number resonances in the discrete $k$-space compared to the continuous one. Given that the system is continuously forced, it will result in an accumulation on the spectrum in the vicinity of the forcing wave number, which in turn will lead to an increase of the nonlinear broadening     $\delta_\omega$. This spectrum accumulation will continue until $\delta_\omega$ reaches the typical values of $\Delta \omega$, at which point the quasi-resonances will become active and will trigger the turbulent cascade in the $k$-space. Thence, turbulent cascade will start to deplete the accumulated spectrum and if the forcing is weak the  broadening $\delta_\omega$ will return to the critical values with $\delta_\omega(k) \sim \Delta \omega(k)$, but not below it because this would switch off the quasi-resonances and re-initiate the spectrum accumulation. Of course, such a sandpile scenario is very speculative and can be only taken as a plausible explanation of the observed behaviour.

\begin{figure*}
	\resizebox{\textwidth}{!}{%
		\includegraphics[scale=0.49]{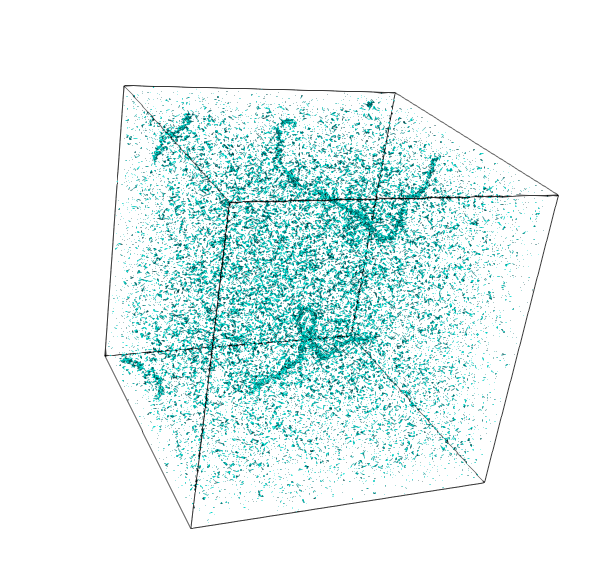}
		\put(-240,10){\bf\large (a)}	
		\includegraphics[scale=0.49]{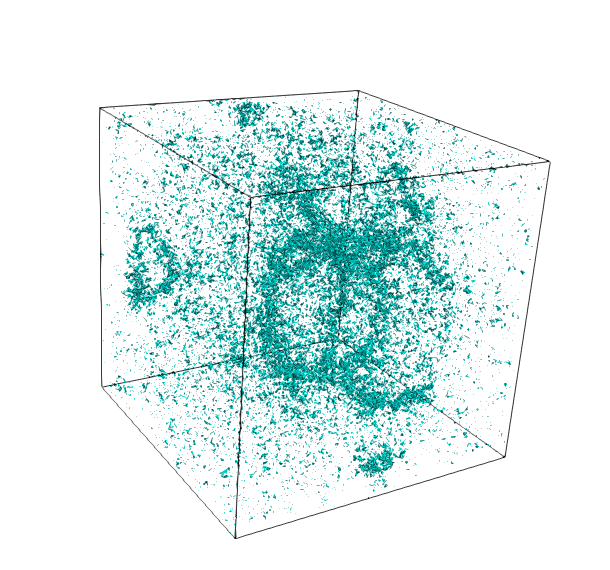}
		\put(-240,10){\bf\large (b)}		
		\includegraphics[scale=0.49]{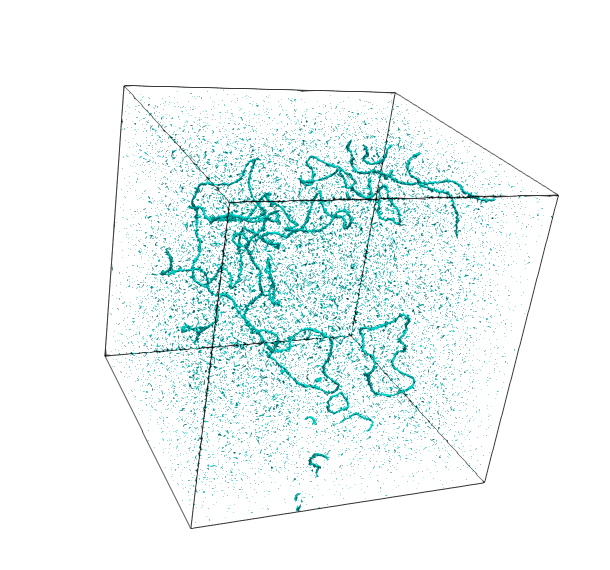}			
		\put(-240,10){\bf\large (c)}	
	}	
	\caption{Isosurface plots of the condensate density for the runs: (a) $\tt Run1$ at $tc^*/\xi^*=8.5\times 10^3$ for $\langle\rho\rangle/\rho^*=0.055$; (b) $\tt Run2$ at $tc^*/\xi^*=6.75\times 10^3$ for $\langle\rho\rangle/\rho^*=0.023$; (c)  $\tt Run3$ at $tc^*/\xi^*= 2.89 \times 10^4$ $\langle\rho\rangle/\rho^*=0.49$.
	}
	\label{iso}
\end{figure*}

We now return to the discussion of the emergence of the condensate as a result of the inverse cascade process. We have already mentioned, while discussing the features of the ST spectra, that the condensate is imperfect: it is nonuniform in the physical space. In Fig.~\ref{iso} (a), (b) and (c) we show the isosurface plots of density $\rho({\r}, t)$ for the runs $\tt Run1$ at $\rho = 0.055 \langle \rho \rangle$, $\tt Run2$ at $\rho = 0.023 \langle \rho \rangle$ and  $\tt Run3$ at $\rho = 0.49 \langle \rho \rangle$, respectively, at times not far from the end of these runs. The chosen threshold values allow us to visualize quantum vortices clearly. We remind that at the vortex center $\rho = 0$ and at distances $\sim \xi$ away from it the density $\rho$ goes back to the values of the surrounding medium (hence the name ``healing length"). In $\tt Run2$ and $\tt Run3$ we see a few well formed vortex lines, which is not surprising, because the condensate fraction reaches rather high values by the end of these simulations (even though the global nonlinearity parameter $\eta $ in $\tt Run2$ remains very low). These vortices are sometimes called phase defects -- they are the physical space manifestations of the condensate imperfection. What is surprising, however, is that the large-scale vortex-like structures are also seen in $\tt Run1$, wherein the condensate fraction remains tiny. These vortex structures are ``fuzzy" or ``bubbly" -- they are surrounded by a great number of small vortex loops, but the overall large-scale coherence of such vortices is evident. Thus, large-scale vortices seem to be robust structures even in cases without well-formed condensate.

Finally, as we had remarked earlier, we are puzzled by the fact that the condensate fraction in $\tt Run2$ grows to a large value whereas it remains tiny in $\tt Run1$, even though the latter is more strongly nonlinear than the former. It seems that the discreteness of the $k$-space, which is a significant factor in $\tt Run2$, helps in the steady growth of the $k=0$ component. A possible reason behind this could that it is harder to ``spill out" particles from the  $k=0$ mode into the neighbouring modes when the dynamics is strongly restricted by the $k$-space discreteness. However, further studies are required to better understand this phenomenon.

\section{Conclusions}
\label{sec:conclusions}

We have elucidated the inverse cascade process that leads to the Bose-Einstein condensation under nonequilibrium conditions within the framework of the force-dissipated GPE. Results obtained from the DNS of the GPE equation in WTT regime are in agreement with the WKE predictions, both the pre-$t_{\star}$ self-similar behavior leading to blow-up in a finite time and appearance of a condensate and the thermalisation process that ensues after the time $t_{\star}$. We find that the characteristics of the inverse cascade regime that develops is strongly dependent on the level of non-linearity present in the system, the latter itself varies with time (as we are not investigating steady states). 

We were able to identify three regimes with different scaling behaviours during the course of the evolution: (i) at early times the presence of $n^{1D}(k) \sim k^{2}$ scaling at wave numbers smaller than the forcing wave number; (ii) followed by the emergence of $k^{-\alpha}$, with $\alpha$ between $0.5$ and $0.8$ for different cases considered here, over wave numbers between $1/\xi$ and the forcing wave number; (iii) finally, $n^{1D}(k)$ tends to develop a plateau, with near zero exponents, at wave numbers close but smaller than the forcing wave number and at still smaller wave numbers we observe a spectrum with large negative exponent ($\sim -2.5$) associated with development of a condensate at $k=0$ mode. Regimes (i) and (ii) correspond to the pre-$t_{\star}$ evolution of the system and at intermediate times these regimes co-exist, whereas regime (iii) is associated with the post-$t_{\star}$ dynamics. Our spatio-temporal analysis clearly shows that if the nonlinearity is small and the finite-box effect is negligible, then the four-wave WTT holds and we obtain $n^{1D}(k) \sim k^{-0.52}$ scaling which is compatible with and corresponds to the WKE predictions of $n_{\omega}(t) \sim \omega^{-x_{\star}}$ with exponent $x_{\star} \approx 1.23-1.24 $. The post-$t^{\star}$ dynamics is dominated, depending on the level of nonlinearity, by weak waves involved in four-wave interactions described by the WTT, or by the presence of coherent structures -- a strong coherent condensate along with well developed loops of hydrodynamic vortices. However, in either of these cases the turbulent spectrum is characterised by an energy equipartition at high wave numbers and a steep spectrum at low wave numbers.

In our simulations, we do not dissipate particles at low wave numbers. The condensate forming at low wave numbers absorbs particles cascading from the high-wave number region of the spectrum.  Na\"ively thinking, eventually we can regard it as an effective particle sink and, therefore, expect the formation of the KZ inverse cascade spectrum with exponent $x=7/6$. However, our numerical simulations show that this is not the case. Indeed, the condensate component cannot serve as an effective sink of particles: it reflects particles back to the high-wave-number range and a spectrum close to the thermodynamic one forms instead of the KZ spectrum. On the other hand, this leaves unanswered the question if the full or a partial suppression of the condensate component (by introducing an extra dissipation mechanism at low wave numbers) could lead to formation of the stationary KZ spectrum. This question should be studied theoretically and numerically in future. In case of the positive answer, one could further investigate if the inverse cascade KZ spectrum could be implemented in a BEC experiment.

\section*{Acknowledgments}
Part of this work was granted access to the computing facilities under GENCI (Grand Equipement National de Calcul Intensif) A0062A10441 (IDRIS, CINES and TGCC) and High Performance Computing Facility at IIT Kharagpur established under National Supercomputing Mission (NSM), Government of India and supported by Centre for Development of Advanced Computing (CDAC), Pune.
V. Shukla acknowledges support from the Start-up Research Grant No. SRG/2020/000993 from SCIENCE \& ENGINEERING RESEARCH BOARD (SERB), India. During this project, S. Nazarenko was supported by the Chaire  D'Excellence IDEX (Initiative of Excellence) awarded by
Universit\'e de la C\^ote d'Azur, France,   Simons  Foundation Collaboration grant Wave Turbulence (Award ID 651471), the  European  Unions  Horizon  2020
research and innovation programme  in the framework of Marie Skodowska-Curie HALT project (grant agreement No 823937) and  the FET Flagships PhoQuS project
(grant agreement No 820392). VS is thankful to Karthik Subramaniam Eswaran for helping with the plots in Fig.~\ref{iso}.

%

\end{document}